# A Two Year Survey for VLF Emission from Fireballs

C. Y. Sung[a], P. Brown[a,b*], R. Marshall[c]

[a]*Dept of Physics and Astronomy, University of Western Ontario, London, Ontario, Canada N6A 3K7*

[b]*Centre for Planetary Science and Exploration, University of Western Ontario, London, Ontario, Canada N6A 5B7*

[c]*Ann and H. J. Smead Department of Aerospace Engineering Sciences, University of Colorado Boulder, Colorado 80303, USA*

## Abstract

Here we report on a two year continuous survey to examine possible VLF signals associated with meteors brighter than magnitude -5. Our survey allowed both calibrated temporal and spatial correlations between VLF signals and fireball lightcurves. We used continuous observations from the Atmospheric Weather Electromagnetic System for Observation Modeling and Education (AWESOME) VLF receiver system (Cohen et al., 2010) deployed at the Elginfield Observatory near London, Ontario, Canada (43N, 81W) to monitor VLF radio signals and correlate with all-sky video recordings of fireballs. This survey from May 2017 to March 2019 was cued using fireballs detected by the Southern Ontario Meteor Network (Brown et al., 2010; Weryk et al., 2007). The GPS conditioned timing of the AWESOME system was continuously synchronized with video recordings directly in the video stream to ensure sub-frame VLF-optical time calibration. The AWESOME system has two orthogonal VLF antennas which permits directional calculation of incoming VLF signals, which were compared to the apparent optically measured locations of simultaneously detected fireballs. VLF events potentially linked to fireballs were checked against the National Lightning Detection Network (NLDN) database to remove false positive association with lightning. During the two year survey interval, over 80 bright meteors (apparent magnitude brighter than -5, brightest recorded event -7.8) were detected and compared to VLF signals detected by the AWESOME system. No definitive evidence was found for VLF emission from meteors up to a limiting magnitude of -7.8.

## Keywords

*Correspondence to: Peter Brown (pbrown@uwo.ca)

## 1. Introduction

Simultaneous sounds associated with the passage of bright fireballs has been reported since antiquity. Unlike delayed sound, understood to be associated with the shock created during the hypersonic flight of a meteoroid, the origin of simultaneous sound is less clear (eg. Romig and Lamar, 1963) .

Two primary hypotheses have been proposed to explain this phenomenon: the electrophonic sound hypothesis and the photoacoustic effect hypothesis. The photoacoustic effect hypothesis proposes that the amplitude of visible light emitted by fireballs fluctuates at the frequency of human hearing and the rapid heating/cooling resulting from this amplitude fluctuation causes air near the material surfaces to vibrate at the same frequency as it heats/cools (Maccabee, 1994). Light curves from bright fireballs have been shown to display such rapid fluctuations (eg. Spurny et al., 2016). Spalding et al. (2017) tested this hypothesis and showed experimentally that the photoacoustic effect can generate noticeable sound when light with suitably modulated amplitude falls on certain materials. This is a promising explanation for anomalous sound, though clearly in need of more experiments and observational verification.

The other explanation offered for simultaneous sound from fireballs is the electrophonic sound hypothesis (Romig and Lamar, 1963). This explanation proposes that electromagnetic (EM) waves produced during the flight of a fireball can induce audible frequency acoustics as the EM wave interacts with conducting objects near observers at frequencies in the range of kHZ to tens of kHz . The transduced VLF radiation results directly in mechanical vibrations of the conducting object as the VLF radiation accelerates charges within the objects and directly creates acoustic waves near an observer. The most convincing validation of this hypothesis, however, would be direct, unambiguous VLF signals detected simultaneously with a bright fireball. While at least one such event has been reported in the literature (Beech et al., 1995) and another proposed (Keay, 1994), similar events have not been convincingly repeated in more recent measurements and so the phenomenon remains controversial.

The study of the electrophonic sound hypothesis started with Nininger (1939)'s proposal of an ether wave related theory, however the existence of ether was quickly disproved. Astapovich (1958), by contrast, concluded that only fireballs that are approximately -9 magnitude have enough energy to produce audible sound through the electrophonic phenomenon. After Astapovich's study, high frequency (MHz to GHz range) EM waves were thought to be responsible for this phenomenon. Hawkin (1958) measured 475, 218, and 30 MHz EM signals in an attempt to directly detect emission from fireballs, but found nothing. He suggested that EM signals of other frequencies might be responsible for simultaneous sounds, but his negative results discouraged further work in this field for two decades.

Keay (1993) examined Hawkin's (1958) work and other literature and discovered that frequencies between 30Hz and 30kHz had not been examined in experiments seeking to detect emission from bright meteors. Keay et al. (1980) suggested that KHz range EM waves are responsible for this phenomenon, caused by direct one to one EM-acoustic wave conversion. Keay (1980) conducted an indoor experiment and subjects successfully perceived acoustic sound converted from KHz range EM waves, experimentally proving the viability of the EM-transduction hypothesis.

There are several indirect sources of evidence which also tend to support this hypothesis. One source of support is that nuclear detonations, whose peak EM intensity is known to be 12kHz, generate "clicks" that are heard at the moment of detonation from a far distance[1]. A similar effect has been documented with lightning (Coleman et al., 2009). In Verveer et al.(2000)'s study of a satellite re-entering the atmosphere they observed both Very Low

---

[1] https://science.nasa.gov/science-news/science-at-nasa/2001/ast26nov_1

Frequency(VLF) electric field signals and audible sounds. Most recently, Obenberger et al (2014) have reported delayed HF and VHF emission from bright meteors detected by the Long Wavelength Array (LWA), proving that some form of high frequency EM radiation is produced by some fireballs. They also stated that the signature they saw from their fireballs suggests fireballs may emit lower frequency, non-thermal pulses.

Keay (1980) proved with human volunteers that electrostatic fields can be heard. Keay (1992) also studied meteor interactions with the atmosphere and concluded a limiting magnitude of -9 is needed to produce a strong enough EM field for observers to perceive electrophonic sound. He also concluded that fireballs lasting longer than 3 seconds are more likely to produce electrophonic sound. Keay and Ceplecha (1994) determined an initial mass of 20kg is needed for electrophonic sound, while also establishing the probable distance range where it could be heard is from 100km up to a maximum of 200 km. They also estimated that one can expect an electrophonic fireball to be heard from any one location on Earth once every 2 or 3 years.

Beech et al.(1995) recorded the VLF signal associated with a Perseid fireball on August 11 1993. The VLF signal was fed into a tape recorder audio track to be synchronized with an allsky video record. A bright Perseid fireball of magnitude -10 recorded with video produced such a strong VLF signal it saturated the spectrogram from 0 up to about 6 kHz, and triggered the recorder's automatic gain control halfway through the fireball. Such a correlated optical fireball and intense VLF emission behaviour has not been reported since.

Garaj et al. (1999) recorded short VLF signal "spikes" that happened during the time of a visible meteor with calibrated VLF/camera timing. The timing of frequently occurring spikes lined up with the peak brightness of the meteor; within the timing uncertainty the optical record and the impulsive VLF signal appear correlated. They concluded that the threshold for VLF emission was around -5 magnitude, however they did not record any electrophonic sounds during the entire Leonid shower in 1998.

Price (2000) claimed to have detected simultaneous VLF signals with fireballs, but the timing coincidence with their optical records was such that clear association was not possible with a particular fireball. Their argument was that the general VLF signals occurred more frequently during the peak of the 1999 Leonid meteor storm. They also argued that the associated VLF signals peaked at a different frequency as compared to lightning associated signals and therefore were from a different source, presumed to be fainter meteors.

Trauter (2002) studied VLF/ELF signals during the 2001 Leonid meteor storm. The VLF associated electric field intensity over long periods of time was integrated and compared to fireball counts during the same period, on a timescale of tens of minutes to hours. While this revealed a probable correlation between meteor activity and the integrated VLF intensity it did not link any single event in which optical fireball light and VLF EM waves were simultaneously detected and analysed at high temporal resolution.

Zgrablic et al. (2002) recorded two cases of electrophonic sounds associated with fireballs produced during the Leonid shower in 1998 having recorded the sound from the fireballs in acoustic microphones that were synchronized to their camera. However their VLF antenna did not detect signals from these events even though their time synchronization was accurate to 0.04s. They explain this by the fact that their VLF receivers were insensitive to frequencies below 500Hz, and the sound they recorded correspond to a 250Hz signal. They concluded that the interaction between fireballs and the atmosphere is stronger than proposed before and noted that the electrophonic signal was not directly emitted by the meteor at its peak brightness. Rather, they think VLF signals are produced before the peak brightness by unknown mechanisms.

Guha et al. (2009) measured VLF signals during the 2007 Geminid shower and found an hourly average of 50 meteors whose VLF signals ranged from 8 to 13 kHz with a mean duration of 6s and a mean bandwidth of 3.6kHz. They did not correlate individual VLF signals with specific meteors, but rather examined the overall spectral distribution of VLF emission which they suggested was correlated with the Geminid shower activity.

Guha et al.(2012) measured the 2009 Leonids shower and claimed to have detected 11,000 meteors per hour in the ELF range below 2kHz. However this number is 137 times more than rate of reported visible meteors. The signal they found peaked at 110Hz, and they describe it as being "very distinct" from background lightning.

Many previous VLF-meteor studies have suffered from either lack of long observing time, or false positive contamination due to lack of common time base between optical and VLF recordings. False positive signals are particularly problematic in VLF measurements as VLF signals are very common in the Earth's atmosphere. Lightning produces up to hundreds of VLF events every minute(Rakov and Uman 2003), and the VLF radio emissions propagate efficiently within the Earth-ionosphere waveguide (Reuveni et al. 2010; Watt 1967). Navies also broadcast strong VLF signals from radio transmitter stations[2]. Naval stations broadcast at specific frequencies continuously, dominating those frequencies and making the identification of  natural signals in certain frequency ranges difficult to isolate.

Since the Earth-Ionosphere waveguide conducts VLF EM waves very well, ambient VLF signals are very abundant. Ambient VLF signals mostly fall under two categories, natural and manmade. Natural ambient signals are dominated by direct lightning signals. Depending on the path the VLF signal took to reach the receiver, its shape in the spectrogram can vary from a straight line to something that looks like an exponential decay curve (whistler); lightning can easily propagate across half the globe to reach an antenna. Man-made VLF sources include

---

harmonics of the 60Hz AC power supply (in North America; 50 Hz elsewhere), as well as the various naval transmission stations whose frequencies are mostly between 11 and 25 kHz[3]. Man-made sources are very obvious in their respective frequencies and tend to be constant in terms of both frequency and broadcast amplitude over long periods of time.

There are various theories as to the underlying mechanism producing electronic sound and hence of VLF emission by fireballs. Keay(1980) proposed that during the heating process of a meteoroid as it enters the atmosphere, the meteor produces ions and electrons which are separated. The build-up of electrons created what Keay described as a "magnetic spaghetti" around the meteor. During the recombination of electron, this "magnetic spaghetti" relaxes and releases its energy as fluctuations in the geomagnetic field in the VLF range. Keay calculated a power requirement of $2*10^{10}$ W at 40km altitude to produce an electric field large enough to be heard at the ground. Keay concluded that this mechanism should only happen for "very large fireballs (-13 magnitude, Hughes 1974)" that can penetrate deep enough into the atmosphere to create a turbulent wake.

Beech and Foschini (1999) proposed a space charge separation model to explain electrophonic emission. In this model, fireball fragmentation creates "bursters", a highly ionized blast wave which propagates into the static geomagnetic field, which then creates a cavity in Earth's magnetic field. A plasma sheet would form around the meteor plasma from easily ionized materials of the ablated meteoroid. If the meteor is large enough, the shockwave would propagate into the plasma. The separation and recombination of electrons at the shock front creates an electric field. According to their model, the threshold for electrophonic sound is $M_v \approx -6.6$.

Kelley et al. (2017) proposed that the head echo plasma of the meteor produces an electric current, and this current in turn generates an electric field antiparallel to the meteor trail.

---

This antiparallel electric field generates a large Hall current in the ionosphere, and the Hall current in turn generates a VLF signal that propagates to the ground. They calculated that an electric field produced by a meteor that has an electron density above $10^{16}\,\mathrm{m^{-3}}$ at its front edge would produce an energy flux above $10^{-8}\,\mathrm{W/m^2}$ at ground level. As the threshold for human hearing is only $10^{-12}\,\mathrm{W/m^2}$, the conversion efficiency for field-to-sound needs to be only 0.1% for the resulting sound to be audible. The electric field strength produced by this proposed mechanism can explain the VLF signal detected by Price et al. (2000). A prediction of this model is that any meteor with dense enough plasma to be detected at GHz frequency by radar as a head echo should be able to produce electrophonic sound audible by the human ear within a range of 100km.

In this paper, we seek to verify the abundance of human hearing frequency range EM signal related to meteors with a VLF antenna system which is spatially and temporally calibrated with optical cameras. We seek to detect signals phenomenologically distinct from typical background VLF noise that we can also correlate in direction and timing t as an observed meteor event. We also explore the possibility of confusion from background VLF signals identified as originating from lightning.

## 2 Experiment

The current study uses a two channel VLF receiver system to record VLF signals on a common time base with a co-located all sky camera at the Elginfield Observatory (43.19279N, 81.31566W) in Ontario, Canada. The all-sky camera correlates events with several other all sky cameras as part of the Southern Ontario Meteor Network (SOMN) providing trajectory and brightness information to a limiting meteor magnitude of -2 (Brown et al., 2010).

The VLF receiver used is the Atmospheric Weather Electromagnetic System for Modeling and Education(AWESOME) system, developed by Stanford University for detection of VLF signals. The AWESOME system consists of two antennas, a pre-amplifier, a line receiver box, a GPS antenna for time calibration, and software pre-written for the system.

This study used a pair of $17.6m^2$ triangular orthogonal antennas for receiving VLF signals. The antennas were set up with the intent to point directly NS and EW. Based on calibration of the antenna system, the NS antenna was actually oriented 4.7±0.3 degrees clockwise from NS and the EW antenna plane was oriented 4.9±0.5 degrees clockwise from EW. Having the antennas oriented orthogonally allows for direction finding based on the ratio of voltages for a common signal. The preamplifier box was placed beside the intersection of the antennas, and a 300 feet coaxial cable runs from the pre-amplifier box back to the line receiver box located inside the Elginfield observatory. The GPS antenna is connected to the line receiver box and is mounted on the roof of the observatory.

Time information is encoded directly to the VLF datastream by the GPS unit and is used to label the collected data. The receiver box then connects to a control computer where the voltage is digitized as 16-bit values from each antenna at 100,000 samples per second using time information from the GPS precise to 100 nanoseconds. The VLF signal is time cropped around events of interest based on automatically recorded bright fireball information from the co-located cameras and stored at a central server.

To correlate VLF signals with fireballs, we use the ASGARD Southern Ontario Meteor Network (SOMN) database of fireball events. This network consists of over 20 all-sky video cameras in the Southern Ontario region to monitor meteor activity (Weryk et al., 2008; Brown et al 2010). The SOMN is sensitive to meteors of apparent magnitude -2 and brighter, corresponding roughly to limiting masses of order tens of grams at low speeds (20 km/s) to as small as 0.1 grams at 60 km/s using the mass-magnitude-speed relation of Jacchia et al. (1967).

Through multi-station cross-referencing of detected video events, the trajectory of meteors is calculated with typical trajectory residuals of order a hundred meters. In particular, the time of each bright meteor and its location are determined from SOMN data. Every morning automated software scans video recordings of the previous night and records every meteor event that occured. A camera located at the Elginfield observatory and hence co-located with the AWESOME system (about 100 meters away from the antennas) is used for direct temporal cuing of the VLF signals. For any meteor event where the Elginfield camera recorded the meteor, the azimuthal information of the fireball direction as measured in the video data is directly compared to the directionality of any signal from the VLF system. For fireballs not recorded by the local Elginfield camera, the meteor path calculated by the ASGARD system is used to calculate direction relative to the AWESOME antennas.

Once fireballs are detected by the ASGARD system, examining the VLF records for possible associated signals requires both spatial and temporal checks to minimize false positive associations. In particular, it is necessary to calibrate the time base between the AWESOME system and the ASGARD system. For the temporal calibration a simple hardware setup was constructed to ensure absolute calibration between the AWESOME VLF timebase and the video time bases. A small LED light was attached to a pole next to the all-sky camera at the Elginfield Observatory. The LED was visible in the field of view of the camera at a fixed location. The LED was connected to a mechanical relay which also was connected to a signal generator which generated VLF signals detectable by the antennas. When the relay was thrown "ON" by the control computer, the LED was switched on and a VLF signal simultaneously transmitted. The VLF signal was detected by the AWESOME system and the on-time of this calibration signal can be isolated to a precision within 0.05 ms. Similarly, the LED was detectable on the video with a precision of one video frame (33 ms).

This calibration was done every hour and was found to generate a very consistent time offset of 1.23 ± 0.03s between the VLF receiver and ASGARD camera. This time offset is only directly measured between the AWESOME system and Elginfield camera. The ASGARD cameras in the SOMN use Network Time Protocol (NTP) for conditioning the computer clocks which in turn time stamp the video. The AWESOME system uses direct GPS timing injected into the signal stream. However, the systematic time offset represents lags present in one or both of the systems. Through comparing meteor lightcurve of the same meteor across different camera stations, the uncertainty in timing between cameras never exceeds two frames. Thus we used a time offset of 1.2±0.1s between the VLF system and cameras elsewhere in the SOMN when only the latter data was available.

For the spatial (or directional) calibration, the relative signal amplitudes between the two roughly orthogonal VLF antennas aligned along the cardinal directions are used (Wood 2004) to provide incoming azimuths for VLF signals. Spatial calibration is performed using the National Lightning Detection Network (NLDN) database. The NLDN data gives the exact time and location of lightning recorded in North America, with timing accuracy on the order of 1 microsecond, and location accuracy of about 400m. Directional calibration involves using a large number of known lightning signals from the NLDN dataset with known arrival azimuths to "fit" the properties of the antennas. In particular, the calibration needs to determine the true angle between the antenna orientations ($\xi$, defined as the difference between the angle made by the antenna and 90 degrees), the azimuthal orientation of the antennas ($\rho$) and the relative gain of each antenna-receiver channel ($\alpha$). To compute the parameters, we use the equation described in Wood (2004):

$\theta true = atan[\alpha*tan(\theta calculated)/cos(\xi) - tan(\xi)\ ] + \rho$

where θtrue is the actual azimuth of the signal. For our calibration, the NLDN direction is considered θtrue. α, ξ, and ρ are the parameters described above. θcalculated is the signal azimuthal direction calculated from the antenna assuming a perfect antenna(i.e. α =1, ξ=ρ=0).

To perform this calibration, four days of NLDN data were used, namely Nov 22, 2017, and June 30, July 5 and July 8th, 2018. The same procedure was used to process data for all 4 days. In principle, each day should produce nearly the same calibration parameters. The variation in these parameters provided a check on the uncertainty in the fit parameters.

Lightning more than 3000km away from Elginfield Observatory were excluded from the calibration as long distance lightning direction can be distorted as they propagate (Said 2009). Lightning events from different locations which overlapped too closely in time (0.03s) were excluded as they interfere with each other and create ambiguous directions.

As a final filter, the R-squared value for the direction calculated across different frequencies was found. Lightning which displayed a significant directional frequency variability (R-squared value lower than 0.8) across different frequencies are also excluded as this suggested the signals from different frequencies may not come from the same source.

Using the known directions for lightning that passed all above filters (both NLDN and calculated from AWESOME data) the final parameters for the calibration of the Elginfield AWESOME antenna were found to be α=1.015±0.020, ξ=-0.5±1.0 degree, and ρ=-4.7±0.5 degrees. The direction comparison between NLDN and our VLF system showed a one sigma spread of 2.5 degrees; this is a measure of our intrinsic directional uncertainty.

*Analysis Methodology*

The AWESOME system began operation in May 2017. Scripts cut ± 30 seconds segments of AWESOME system data whenever the ASGARD system detects a meteor. Data around the time of meteors and calibration related data are both periodically synchronized to another server. We thus have a database of 80 1-minute long segments of VLF data centered around time of bright meteors detected by the ASGARD system for this two year interval.

To determine the true incoming direction of short duration VLF signals, amplitude ratio methods were adopted from Said (2009). For a signal that has a short duration but spans a large range of frequencies, the raw data is first inspected for the duration of the signal of interest. The signal is usually apparent in both raw and filtered amplitude, thus it is easy to manually determine the optimal Fourier transform window to use to analyze the data by simply examining the raw/filtered amplitude around the region of the signal. After the transformation, the amplitude of all frequency ranges over the time window containing the signal are extracted. This is equivalent to taking a vertical cut through the spectrogram, and looking at its frequency versus amplitude.

The extracted NS and EW amplitudes are compared against each other. If a signal is indeed from one source, the ratio between its north-south and east-west channel amplitudes will be very consistent across all frequency ranges. The resulting plot of NS versus EW amplitude will show a very tight line showing the direction of the incoming signal. The raw amplitude ratio gives four possible directions, but the phase difference between the two channels can be used to reduce the possibilities down to two. For our antenna system, signals are in-phase for northwest and southeast signal sources, and out-of-phase for northeast and southwest signal sources. Thus a 180 degree azimuthal ambiguity is always present; however, for most events this is adequate for determining the relationship between any observed meteor candidate and a recorded VLF signals.

Since the human hearing range goes from 20Hz to 20kHz (Rosen, 2011), and the electrophonic sound model assumes a 1-to-1 conversion of VLF fireball emission to acoustic waves, we assume any potential fireball produced VLF signal will lie between 100Hz and 20kHz and ignore lower frequencies. VLF produced signals from nuclear weapon detonations which peak at 12kHz[4] and Beech et al.(1995)'s study of a saturated VLF signal associated with a bright Perseid fireball extending up to 10kHz support this choice.

## 3. Results

### *3.1 Event Selection*

From the list of meteors detected by ASGARD from May 2017 to March 2019, we select events that are brighter than -5 magnitude and are within 150km radius of the VLF system to further inspect. The magnitude limit is set by predictions from various studies which suggest events fainter than -5 will not produce VLF emissions. Over the past few decades, different studies produced a range of estimates as to what fireball magnitude is needed to emit detectable VLF radiation. The estimates range from -6 (Romig and Lamar, 1963) to -13 (Hughes 1974). On this basis, a -5 threshold was selected for our study.

Fireballs not meeting the distance filter but that are exceptionally bright are also individually inspected (eg. the 2018 Hamburg Fireball - Brown et al (2019)). The distance filter is used as a simple estimate for events expected to be clearly visible in the VLF system to avoid potentially weak signals easily confused with background noise.

---

[4] https://science.nasa.gov/science-news/science-at-nasa/2001/ast26nov_1

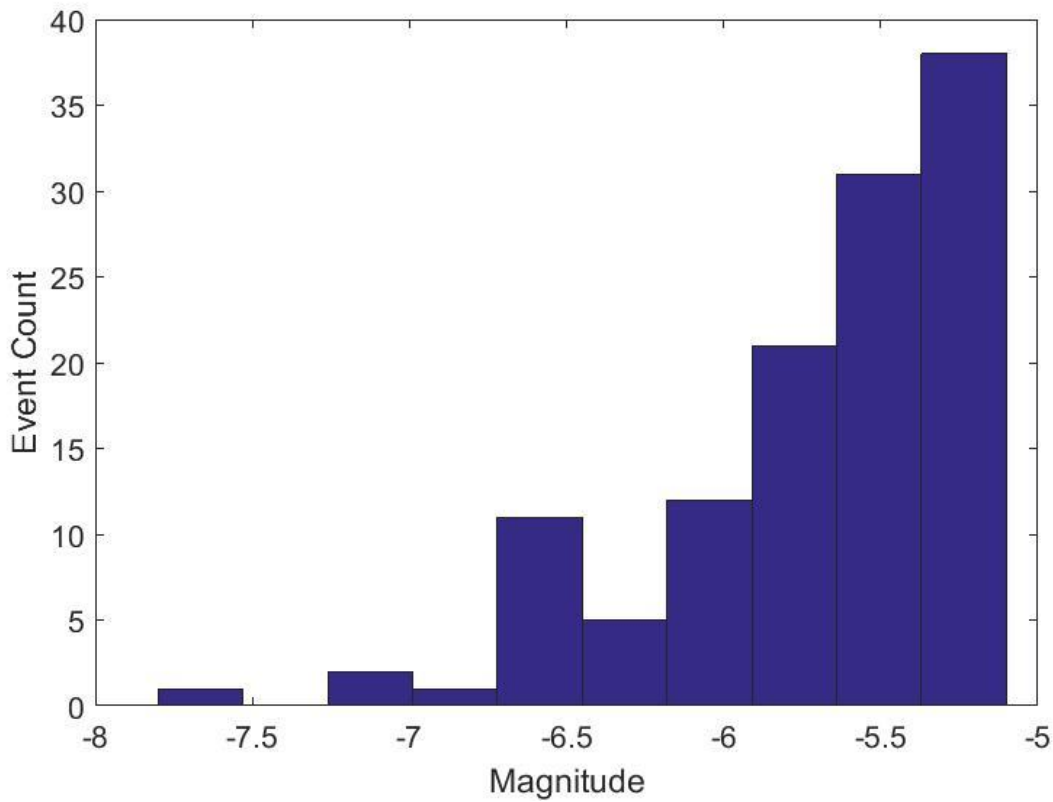

Figure 1. Magnitude distribution of the ASGARD fireball events selected for our survey. Only three meteors exceeded a magnitude of -7. The majority of selected events have a magnitude dimmer than even the most conservative theoretical estimate of the limit of VLF production of -6.

### 3.2 Event analysis

Over the duration of this survey, 80 fireballs were recorded by the ASGARD system which met our filter criteria. These events also had VLF AWESOME data recorded. To manually inspect these events, all were plotted as time matching lightcurve-spectrogram pairs to visually scan for any signal that is not similar to the lightning signals observed as shown in figure 2. A typical lightning duration of 0.6ms (as observed when doing antenna direction calibration) is in sharp contrast to the many seconds long saturated signal of the one probable fireball VLF signal Beech et al. (1995) reported. Among the 80 events, none showed anything that is clearly phenomenologically different from a typical VLF lightning signal. No signal lasted longer than a few hundred milliseconds. However, 70 events showed some kind of temporally correlated VLF signal within the duration of the light curve.

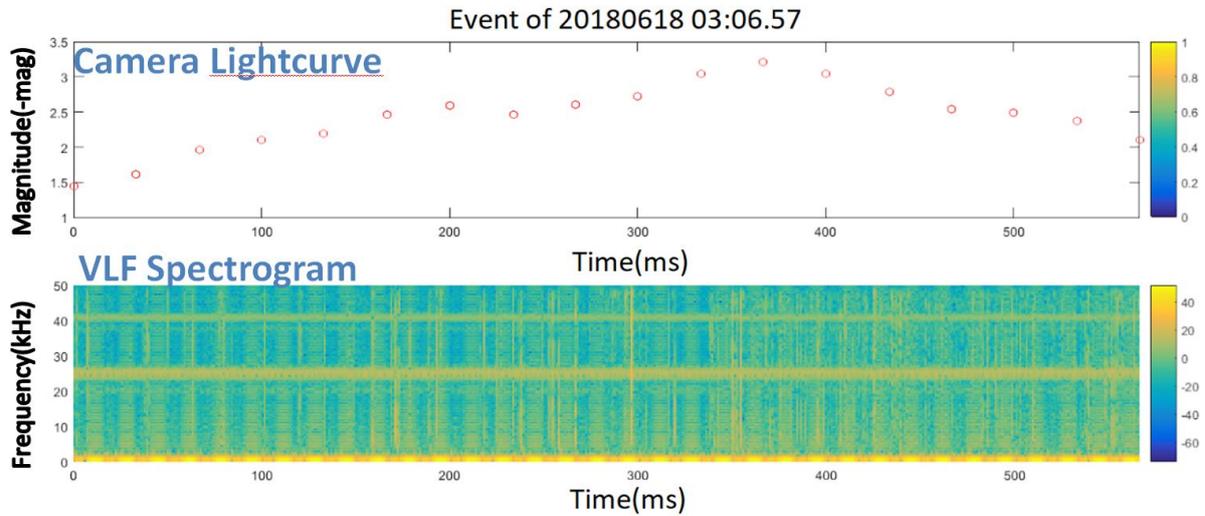

Figure 2. A Light curve-spectrogram comparison. Top is the light curve of meteor event 2018/06/18 03:06.57, each point corresponds to the magnitude of the meteor calculated from one frame camera recording. Bottom is the spectrogram of NS channel at the time of the light curve. The offset between the light curve and the spectrogram is calibrated to within one frame of uncertainty(33ms).

The VLF signal during the time of each of these 80 events was Butterworth filtered to remove unwanted frequencies (frequencies above 20kHz outside human hearing range). Then they were transformed through a script(Cohen et al. 2010) to remove human power line harmonics that showed up very strongly in lower frequency bins of the spectrograms. Filtered data was first examined for any long duration (resembles the spectrogram of the event reported by Beech et al.(1995)) signals; none were found. Filtered data were then manually examined for any outstanding peaks in signal amplitude. The direction of each individual peak was then calculated.

We found that there are typically a few to a few dozen VLF impulsive signals during any one meteor lightcurve, depending on the length of the light curve. All VLF peaks are fairly short: the longest ones are only a few milliseconds, nowhere near the long duration high intensity signal recorded by Beech et al. in 1995, the single published example of a probable VLF signal from a bright fireball. The peaks thus can be treated like lightning (which most probably are), and we use the amplitude ratio across all frequencies to compute an incoming signal direction. The same process used to analyze the lightning signals for calibration is used on these peaks observed to occur during the meteor. Most showed a very tight fit, indicating all frequencies come from the same signal source, with a few exceptions displaying two clear sources across different frequencies. Of the 80 events showing VLF signals coincident in time, only 13 showed one or more of the time-correlated peaks which occurred during the duration of the light curve also having a calculated direction that matched the range of meteor azimuths as measured by the ASGARD system.

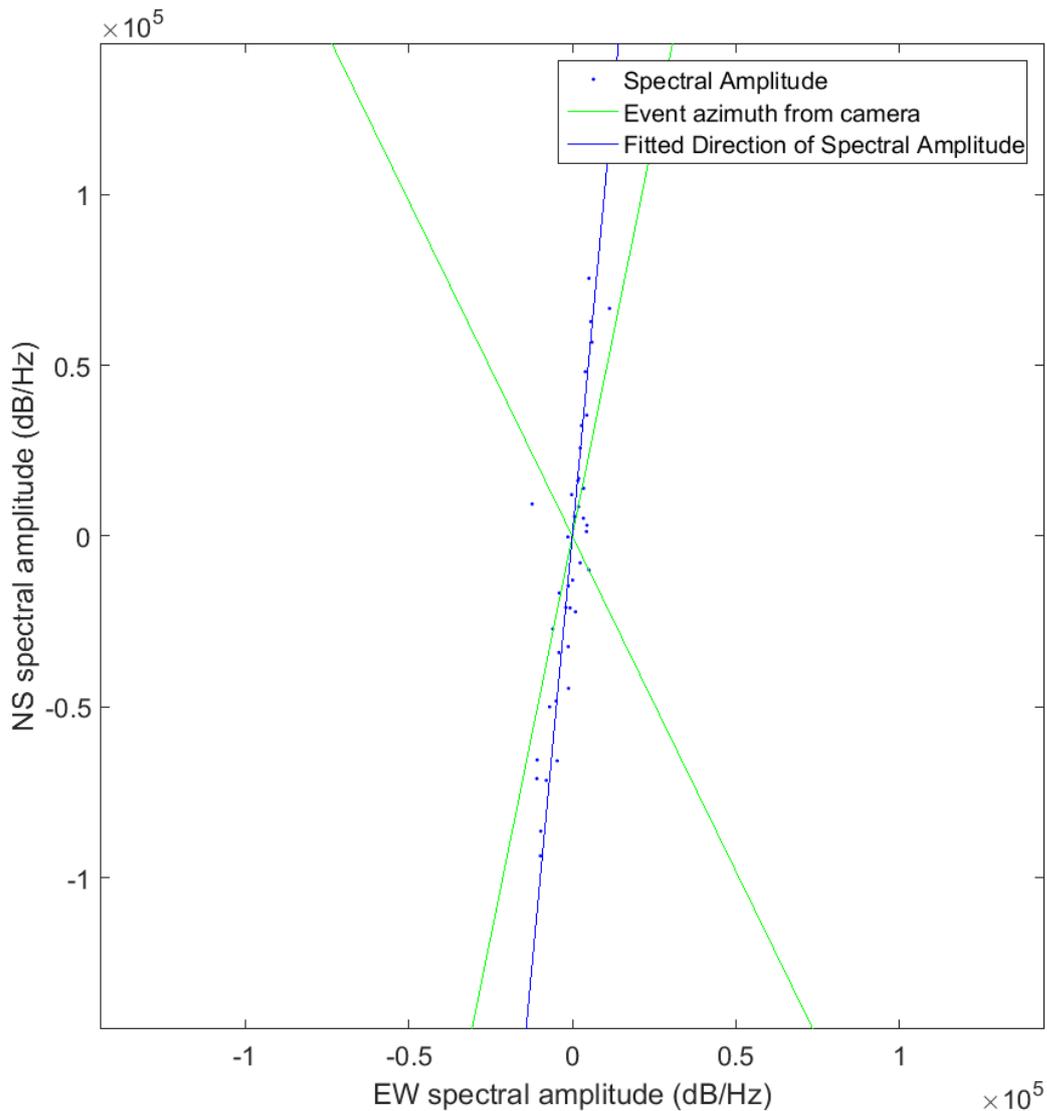

Figure 3. Incoming Azimuth of one of the many VLF impulsive signal peaks within the duration of fireball event 2017/08/23 06.25.22. Shown are the relative NS vs EW spectral amplitudes (y and x axis respectively). Each blue dot is the NS/EW amplitude in one frequency bin of ~500Hz width, starting at 3kHz and extending to 20 kHz. The blue line is the line of best fit of the blue dots, which is taken as the incoming azimuth of the VLF signal. The green lines are the starting and ending azimuth of this particular meteor as seen on the ASGARD camera at the Elginfield Observatory. Most peaks during this light curve do not match the direction of the meteor and is not shown in this fit.

*3.3 NLDN Lightning Event Filter*

Since all 13 VLF signals concordant in time and having arrival directions coincident with the observed fireball showed spectra typical of lightning, a simple explanation is that these are in fact unrelated lightning. To check this hypothesis further, NLDN data were requested for the dates/times of all 13 events to verify whether a given associated signal is indeed lighting that happens to match in both time and direction a particular fireball. This is done by simply looking at the NLDN database around the time of those 13 events, and computing the equivalent azimuth of any lightning that match in time. Of the 13 fireball events of interest, corresponding lightning (having time and direction matches) were found for 6 of them. The remaining 7 were not correlated with any lightning within North America. However, as the NLDN dataset is not global in nature we cannot rule out these signals as being from more distant lightning not in the NLDN database, or from a stroke missed by NLDN, which has a detection efficiency of 95% for cloud-to-ground lightnings and only 60% for intra-cloud lightnings within the USA[5,6]. However intra-cloud lightning is about 3 times more prevalent than cloud-to-ground lightning. Indeed, all 7 of these fireballs that had time and direction correlated with an impulsive VLF signal but which could not be positively linked to an NLDN lightning event also had additional VLF signals which occurred during the time window of the meteor, and none of the additional signals were a directional match to the meteor.

---

[5] https://www.vaisala.com/sites/default/files/documents/MET-G-NLDN-Brochure-B210412EN-E_Low.pdf

[6] https://ghrc.nsstc.nasa.gov/uso/ds_docs/vaiconus/vaiconus_dataset.html

*3.4 NLDN Backward Check*

　　To better assess the completeness of the NLDN events relative to the signals detected by the VLF system, a backward check was performed to investigate the percentage of lightning visible on our VLF spectrogram that was also reported by the NLDN system. This provides an estimate of the fraction of global lightning detected by our VLF system as opposed to North America-only lightning events, reported by the NLDN. We find an estimated 4-5 lightnings per second as the average observed rate with the VLF system. Therefore it is necessary to know if a result of "no corresponding lightning was found in NLDN database" implies a high probability that the signal of interest is not really lightning.

　　To determine this quantity, 5 random segments of 3 second intervals were chosen from VLF system data and lightning recorded by NLDN were compared. At best we found that NLDN records a quarter of what can be seen on the AWESOME system; that is only one quarter of impulsive VLF signals detected by AWESOME are correlated with lightning in the NLDN. In one of the examined time segments, NLDN reported no lightning events in North America for $\pm 1$ seconds of the selected time frame, while tens of lightning-like signals were visible in the AWESOME system spectrogram. So the correlation with NLDN means that the possibility of the suspected VLF signals being lightning cannot be very effectively ruled out by their absence in the NLDN database.

*3.5 VLF - fireball correlated events*

　　There are a total of 13 fireballs during our survey which showed VLF signals which occur during the luminous time of the meteor and have at least one impulsive signal with an arrival direction consistent with the direction of the meteor as seen from Elginfield. Seven of these events had no corresponding lightning record in the NLDN database. The range of observed fireball azimuthal spreads varies from 1 degree to 48 degrees. All associated fireballs have multiple VLF peaks during the time of the meteors luminous flight,but very few of the peaks match the meteor apparent azimuth as seen from the Elginfield Observatory. The peaks which do show a spatial correlation are mostly not the strongest peak. The strongest peaks of most events point in directions not corresponding to the meteor. Since only bright fireballs events at close range are selected for inspection, if they do indeed produce VLF signals, we expect that their signal should dominate the spectrum and be very distinguishable from distant lightning. Details of all seven of these events are given in the Supplementary Material, and raw data uploaded to Mendeley online database.

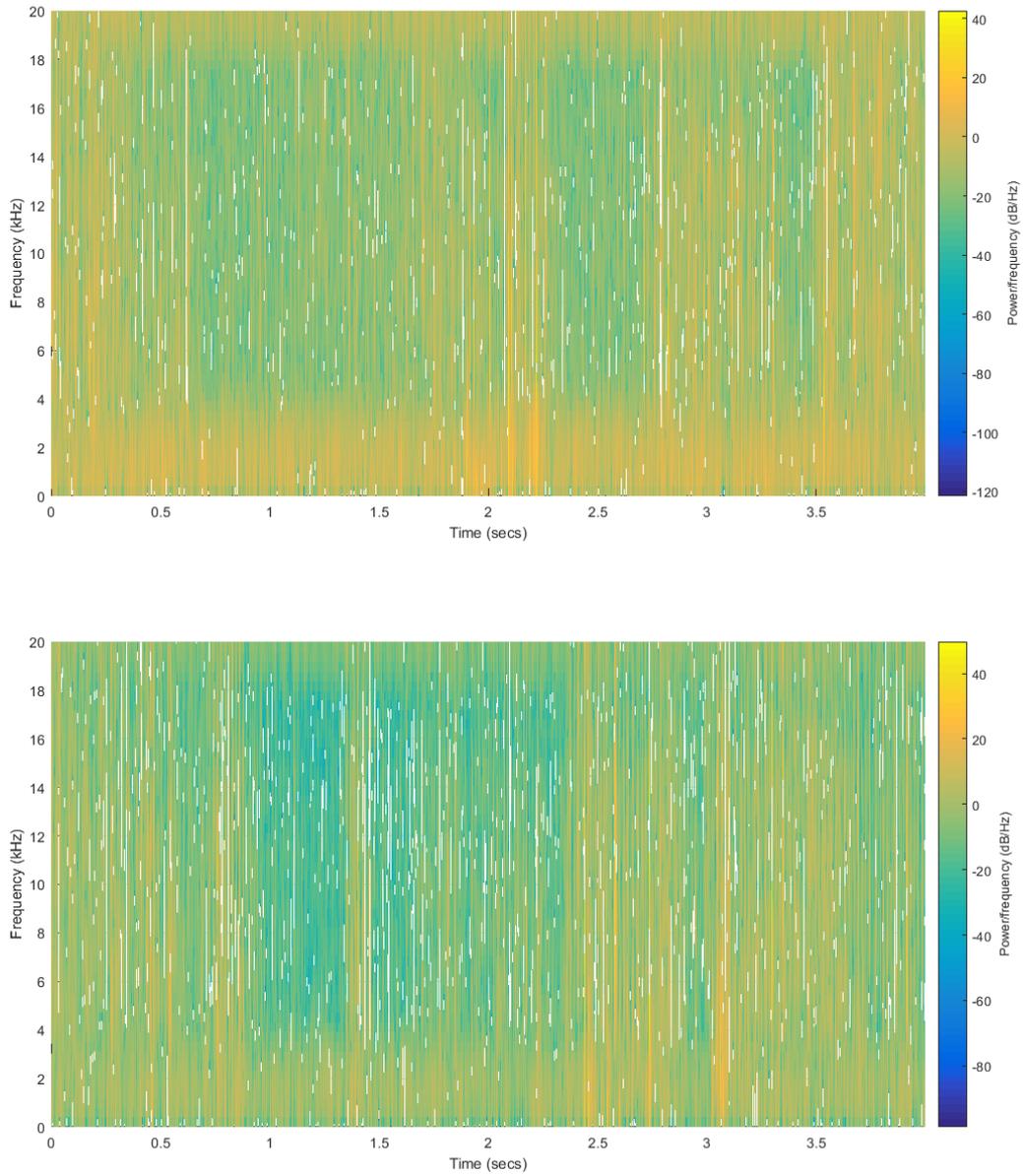

Figure 4. Top: ±2 second spectrogram of a lightning detected by the NLDN. Bottom: ±2 second spectrogram centered around the light curve of meteor 2018/08/15 08:56.28. They do not have any significant difference in terms of phenomenology.

## 4. Discussion

### *4.1 Chances of random signals passing the filters*

#### *4.1.1 Time*

According to Christian et al. (2013), there are 44±5 lightning strikes per second anywhere on earth. The duration of a typical ASGARD-detected meteor lightcurve ranges from 0.2 second to 1 second, with exceptionally long ones being up to 3.6 seconds. Assuming Poisson statistics for these random events, $p(x; \mu) = \frac{\mu^x}{x!}e^{-\mu}$, where x is the number of events, and μ is the average amount of events for the duration, we obtain the probability for no lightning strikes to happen anywhere on Earth within 0.2, 1, and 3.6 seconds to be

$$p(0; 44 * 0.2) = \frac{\mu^0}{0!}e^{-8.8} = e^{-8.8} = 1.5 * 10^{-4}$$

$$p(0; 44 * 1) = \frac{\mu^0}{0!}e^{-44} = e^{-44} = 7.8 * 10^{-20}$$

$$p(0; 44 * 3.6) = \frac{\mu^0}{0!}e^{-158.4} = e^{-158.4} = 1.6 * 10^{-69}$$

The latter two have very little meaning aside from saying it is next to impossible to not have a lightning strike somewhere on earth for the duration of a lightcurve that is more than 1 second long. However, not all lightning events produce a detectable signal at our VLF system, hence the calculated probability above is an upper limit to what is actually observed. Although there is close to a 100% chance that at least one lightning event will occur during each meteor event, not all 80 meteor events in the survey showed at least one lightning recorded within its duration, demonstrating only that the chance of temporally correlated false positives is very high. Without detailed knowledge or modelling of the propagation of VLF from lightning to our particular receiver we can only assert the above values as extreme upper limits. In fact, we find from our examination of random time windows that typically AWESOME detects 4.5 impulsive, lightning-like events per second, making the above numbers change to 4% chance of no lightning detected by the AWESOME system for the average meteor duration of 0.739s among our dataset of 80 meteors.

$$p(0; 4.5 * 0.2) = \frac{\mu^0}{0!} e^{-.9} = e^{-0.9} = 0.41$$

$$p(0; 4.5 * 0.739) = \frac{\mu^0}{0!} e^{-4.5*0.739} = e^{-3.33} = 0.04$$

$$p(0; 4.5 * 1) = \frac{\mu^0}{0!} e^{-4.5} = e^{-4.5} = 1.1 * 10^{-2}$$

$$p(0; 4.5 * 3.6) = \frac{\mu^0}{0!} e^{-4.5*3.6} = e^{-16.2} = 9.2 * 10^{-8}$$

The actual rate of finding at least one lightning signal in a meteor lightcurve time window is 70/80=87.5%, slightly lower than the estimate of 96% from this calculation but reasonably consistent. This can be explained by the variation in duration of the meteor luminous flight time and the fact that the VLF impulsive signal rate varies diurnally. Long duration (more than 1 second long) meteors with multiple lightning correlations only count as one detection, but they still increase the average duration of meteors.

### 4.1.2 Direction

For the dataset in this survey, the list of 80 meteor events has an average azimuthal spread of 12.03 degrees, with a standard deviation of 11.98 degrees. Azimuthal spread is defined here as the difference between the azimuth of the observed meteor at the start and the end of its luminous flight as seen from the Elginfield Observatory.

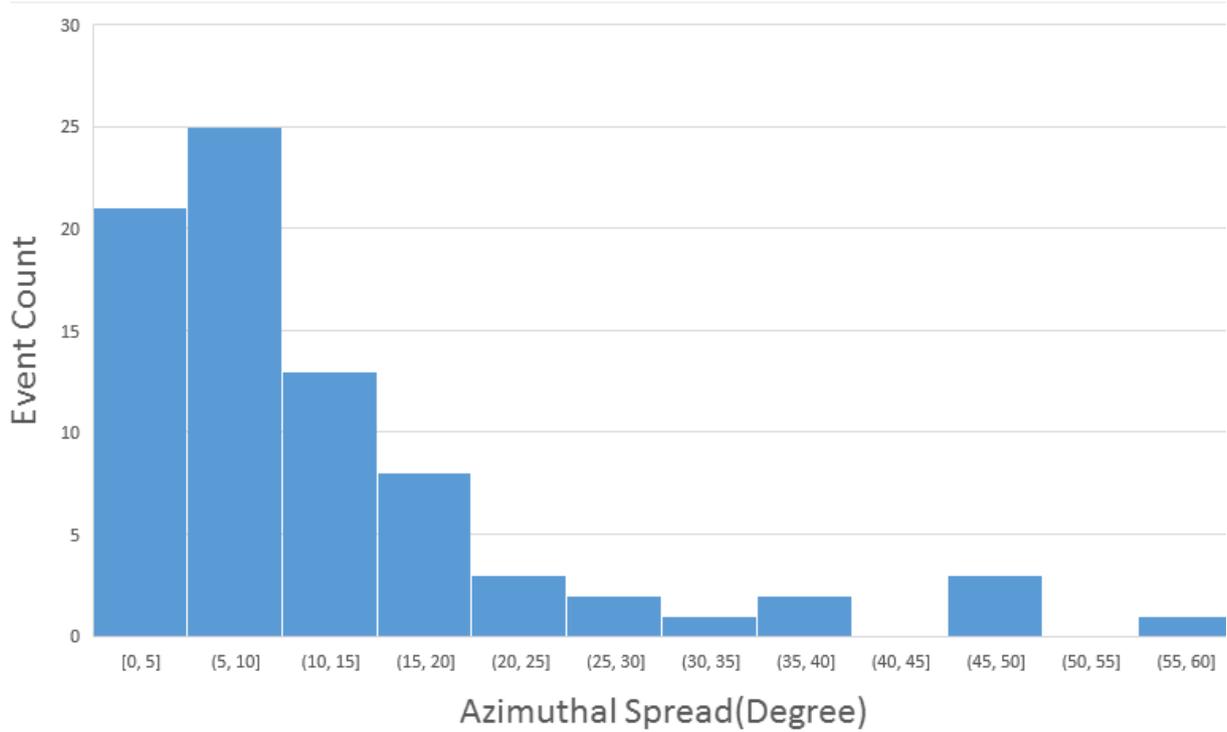

Figure 5. Azimuthal spread distribution of the selected events. Most events traverse an angular path smaller than 15 degrees.

Of the total 80 meteors, 21 made an angular path smaller than 5 degrees, 25 between 5 and 10 degrees, 13 between 10 and 15 degrees, and only 21 more than 15 degrees. For directional false positives, the probability of lightning falling in the range of the meteor is just the azimuthal spread of the meteor divided by the 180 degrees of possible incoming azimuth, assuming for simplicity that lightning occurs equally probable in all directions. This is likely a reasonable assumption if we assume global detection. For a meteor event that ranges 5 degrees, the probability is 3%. The average of 12 degrees implies that on average there is a 7% chance for any random lightning to appear to correlate within the azimuthal spread of an average meteor trajectory in our ASGARD dataset.

### 4.1.3 Overall rate

For the upper limit, for an average duration meteor, the chance of lightning happening in this window is close to 100%, the rate of lightning happening at the same time and within the azimuthal spread of the meteor is almost the same as the probability of its direction being correct. Using the binomial distribution, for a per event chance of 7%, the number of events needed for a certain chance of happening is $f(p, n, k) = \frac{n! p^k (1-p)^{n-k}}{k!(n-k)!}$, where p is the probability, k is the number of success, and n the number of trials. For this case,

$$f(0.0668, n, 0) = \frac{n! \, 0.0668^0 (1 - 0.0668)^n}{0! \, n!} = (1 - 0.0668)^n$$

The average duration of a meteor event for this sample is 0.739 seconds,

$$f(0.0668, 0.739 * 44, 0) = (1 - 0.0668)^{0.739*44} = 10.6\%$$

Using our average meteor duration of 0.739 seconds, there is thus an 89.4% chance of lightning happening during the window of its luminous flight somewhere on Earth (assuming a global rate of 44 lightning strikes per second) and also having the same azimuth as the meteor as observed from the Elginfield Observatory.

However for the actual observed rate of 4.5 lightning-like impulsive events detected by our VLF system per second,

$$f(0.0668, 0.739 * 4.5, 0) = (1 - 0.0668)^{0.739*4.5} = 79.5\%$$

Thus, there is a 20.5% chance that the VLF system will detect a lightning signal during the time of luminous flight of a meteor where that signal also has an arrival direction within the azimuthal spread of the meteor. The actual detection ratio is 7/80=8.75%, roughly half of our theoretical estimate from this simple calculation. This is remarkably similar and the factor of two difference can be explained by the variation in duration of the meteor luminous flight time and the non-isotropic nature of actual lightning azimuth.

*4.2 Characteristics of VLF signals for matching cases*

After calculating the likelihood of a signal passing both filters mentioned above, we examined the spectrogram of the seven remaining individual events. For all meteor events where a peak matches both time and direction, there are always more peaks in the same timeframe whose direction does not match the meteor.

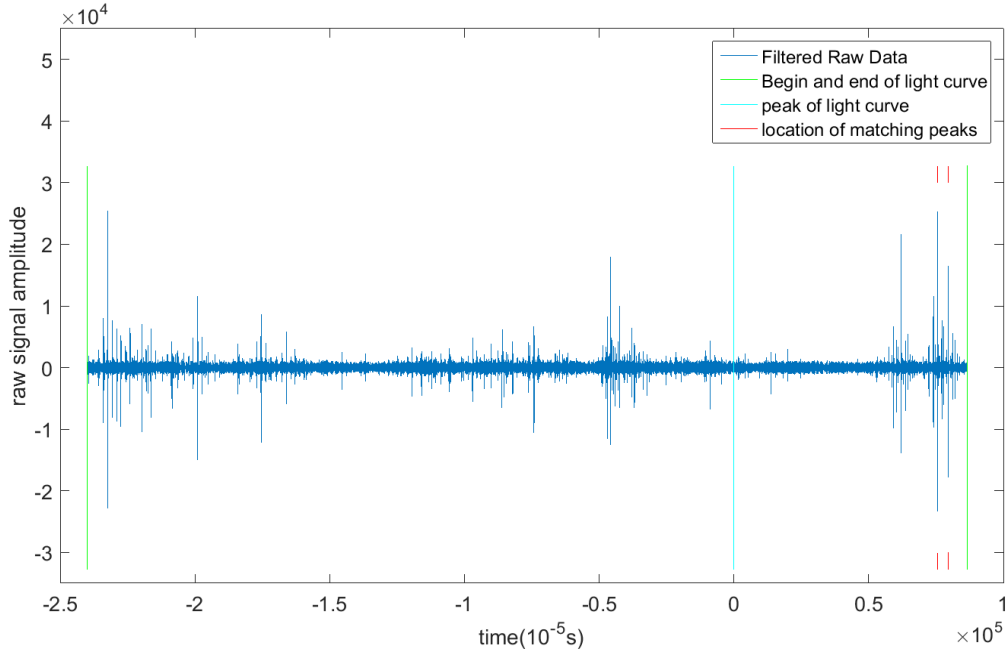

Figure 6. Location of the peaks that matches meteor direction for event 2017/08/23 06.25.22. All other peaks of similar or larger magnitude within the duration of the lightcurve does not match the direction of the meteor.

None of the peaks that match both time and direction dominate their spectrum; they are always one of the many indistinguishable signals in the duration of meteor light curve. Moreover, they show no significant spectral difference from lightning. None of our possible signals shows the sort of long duration, strong spectral peaks and behaviour of the one probable VLF-fireball signal reported by Beech et al (1995). This suggests that the peaks that passed the above two filters are the same as their neighbours. The simplest explanation is that they are just lightning that happened at the time of the meteor and from the same direction. They are not direct VLF emission from meteors and in fact we find no evidence in our two year survey of any credible VLF emission from any bright meteors. However, we note that the number of very bright meteors in our survey is quite small; none were comparable to the energy of the Beech et al (1995) fireball, for example.

**5 Conclusions**

We performed a two year survey to search for potential electromagnetic emission from meteors in the VLF range. We based our selection of the brightness of examined events on past theoretical models and assumed a one-to-one conversion of VLF signal to audible sound. We use the ASGARD system operating in the SOMN to obtain meteor trajectory and energy information in the region around our VLF detector. For a VLF detector we used the AWESOME system developed by Stanford University. The AWESOME system was setup at the Elginfield Observatory located in southern Ontario, within the area covered by the Southern Ontario Meteor Network.

Unlike previous studies where precise time and direction information was not available due to equipment limitations, our VLF system was temporally linked directly to a co-located all sky camera allowing for rigorous calibration of detected meteors in both time and direction. Precise time and direction calibrations was performed between the VLF system and ASGARD system in an effort to limit false positives being identified as VLF signals from meteors. Timing was calibrated to a precision of 0.03s, and direction was calibrated to a precision of 2.5 degrees.

In total, 80 fireballs were inspected for the survey. Of these, 70 showed VLF signals within the time window of the meteor luminous flight, and 13 showed directionally matching VLF signals within the meteor duration window. North American corresponding lightning was only found in the NLDN database for 6 out of these 13 events. Seven events had VLF signals within the duration of the meteor lightcurve that matched the direction of the meteor and also had no corresponding lightning record in the NLDN database. However, all seven events showed spectra similar to lightning and had other VLF signals within the meteor lightcurve time window that did not match the meteor direction. All seven of these possible VLF correlated signals had waveforms, duration, and frequency range indistinguishable from typical lightning signals. The seven events also showed no trend in magnitude, showing no correlation between meteor magnitude and directionally matching signals. We would expect a higher likelihood of real VLF signals from brighter (more energetic) fireballs as predicted by all existing theories of meteor-VLF emission, but did not see any such correlation.

As most of our fireball events were between -5 and -7 apparent magnitude, on the basis of non-detection of a credible VLF signal from any meteor over the course of the two year survey(a total of 77 magnitude -5 to -7 events), we can conclude that VLF signals from fireballs are a very rare occurrence for fireballs dimmer than magnitude -7.

For meteors brighter than -7 magnitude, we do not have a significant sample over the course of the two year survey to make strong conclusions as to the rate of VLF emissions. We can only say that nothing was detected from the 3 events over the past 2 years that were brighter than -7 magnitude. Clearly a longer survey is required, focusing on very bright events, to rule out common VLF emission from bright fireballs.

**References**

Beech M. et al, 1995, VLF Detection of Fireballs. Earth, Moon, and Planets 68, 181-188

Beech M., Foschini L., 1999, A space charge model for electrophonic bursters. A&A, 345, L27.

Beech M., Foschini L., 2001, Leonid electrophonic bursters. A&A, 367, 1056-1060, doi:10.1051/0004-6361:20000436.

Brown P.G. et al, 2019, The Hamburg Meteorite Fall: Fireball trajectory, orbit and dynamics., Meteoritics and Planetary Science, doi:10.1111/maps.13368.

Christian H.J. et al., 2003, global frequency and distribution of lightning as observed from space by the optical transient detector, j. geophys. res., 108, d1, acl 4-1, doi:10.1029/2002jd002347

Cohen, M. B., et al., 2010, Mitigation of 50-60 Hz Power Line Interference in Geophysical Data., Radio Science, 45, 6, doi:10.1029/2010rs004420

Coleman T. a., Knupp K. R., and Tarvin J. T., 2009, Review and Case Study of Sounds Associated with the Lightning Electromagnetic Pulse. Monthly Weather Review 137:3129–3136. doi:10.1175/2009MWR2931.1

Guha, A., De, B.K., Choudhury, A., & Roy, R., 2012, Spectral character of VLF sferics propagating inside the Earth-ionosphere waveguide during two recent solar eclipses, Astrophysics and Space Science, 341, 287. doi:10.1029/2011JA017498

Hughes D. W., 1974, The Influx of Visual Sporadic Meteors, Mon. Not. R. Astron. Soc., Vol 166, p. 339. doi:/10.1093/mnras/166.2.339

Jacchia, L., Verniani, F., & Briggs, R. E, 1967, An Analysis of the Atmospheric Trajectories of 413 Precisely Reduced Photographic Meteors, SCoA, 10, p1-139. doi:10.5479/si.00810231.10-1.1

Keay C. S. L. , 1980, Anomalous Sounds from the Entry of Meteor Fireballs, Science, New Series, 210, 4465, 11-15, doi:10.1126/science.210.4465.11

Keay C. S. L., 1992, Electrophonic sounds from large meteor fireballs, Meteoritics 27,144-148. doi:10.1111/j.1945-5100.1992.tb00741.x

Keay C. S. L. , Ceplecha Z., 1994, Rate of observation of electrophonic meteor fireballs, J. Geophys. Res., 99, E6, 13,163-13,165. doi:10.1029/94JE01092

Kelley M.C., Price C., 2017, On the electrophonic generation of audio frequency sound by meteors, GeoRL, 44, 2987. doi:0.1002/2017GL072911

Lamar D. L. , M. F. Romig, 1964, Anomalous Sounds and Electromagnetic Effects Associated with Fireball Entry, Meteoritics, Vol2, No.2. doi:10.1111/j.1945-5100.1964.tb01419.x

Obenberger K. S. et al. 2014. Detection of radio emission from fireballs. Astrophysical Journal Letters 788:2–7. doi;10.1088/2041-8205/788/2/L26

Price C., Blum M., 2000, ELF/VLF radiation produced by the 1999 Leonid meteors, Earth, Moon and Planets 82-83, 545-554. doi:10.1023/A:1017033320782.

Reuveni Y. et al., 2010, Natural atmospheric noise statistics from VLF measurements in the eastern Mediterranean, Radio Science, Vol45, RS5015. doi:10.1029/2009RS004336, 2010

Rosen S., 2011, Signals and Systems for Speech and Hearing (2nd ed.). BRILL. p. 163.

Rakov, V.A. and Uman, M.A., 2003, Lightning: Physics and Effects. Cambridge University, Cambridge, p. 687. doi;10.1017/CBO9781107340886

Said R.K., 2009, Accurate and Efficient Long-Range Lightning Geo-Location Using a VLF Radio Atmospheric Waveform Bank, Stanford University

Silber E.A., 2014, Observational and Theoretical Investigation of Cylindrical Line Source Blast Theory Using Meteors, University of Western Ontario - Electronic Thesis and Dissertation Repository. Paper 2112

Spalding R. et al, 2017, Photoacoustic Sounds from Meteors, Sci.Rep, 7:41251. doi:10.1038/srep41251

Spurny P. et al, 2017, Atmospheric trajectory and heliocentric orbit of the Ejby meteorite fall in Denmark on February 6, 2016, Planet. Space Sci. Vol 143, 192-198

Trautner R. et al., 2002, ULF-VLF Electric Field Measurements During The 2001 Leonid Storm, Proceedings of Asteroids, Comets, Meteors - ACM 2002, International Conference, 29 July - 2 August 2002, Berlin, Germany.

Verveer A. et al., 2000, Electrophonic Sounds from the Reentry of the Molniya 1-67 Satellite over Australia: Confirmation of the Electromagnetic Link, Meteoritics and Planetary Science Supplement. 35. 163.

Watt A. D., 1967, VLF Radio Engineering, Pergamon Press

Wood T.G., 2004, Geo-Location of Individual Lightning Discharges Using Impulsive VLF Electromagnetic Waveforms, Stanford University

Zoghzoghy F.G., 2015, Statistical Analysis and Modeling of Lightning Using Radio Remote Sensing, Standard University

Web References

https://science.nasa.gov/science-news/science-at-nasa/2001/ast26nov_1

https://ghrc.nsstc.nasa.gov/uso/ds_docs/vaiconus/vaiconus_dataset.html

http://solar-center.stanford.edu/SID/AWESOME/ScienceIntro.pdf

https://vlf.stanford.edu/research/introduction-vlf

https://vlf.stanford.edu/research/subsurface-detection

https://www.vaisala.com/sites/default/files/documents/MET-G-NLDN-Brochure-B210412EN-E_Low.pdf

https://www.mwlist.org/vlf.php

https://sidstation.loudet.org/stations-list-en.xhtml

**Supplementary Material. Seven meteors where matching lightning VLF signals were not found**

This section contains the plots for the seven events where VLF impulsive peaks matched both the time and the direction of ASGARD-detected meteors. The raw data for all these events are uploaded to mendeley database.

In the following plots for each case, the first plot(.1) is the VLF signal time versus NS VLF amplitude (filtered to remove powerline harmonics using the technique of Cohen et al.(2010)). This plot shows the timing of the VLF peak(s) within the duration of meteor luminous flight. The green vertical lines mark the beginning and end of the optical signal of the meteor detected by the ASGARD camera while the blue vertical line marks the time of maximum brightness of the meteor. The timing of VLF impulsive signals having azimuths within the observed range of meteor azimuths are shown with short vertical red lines above and below the peak(s).

The second plots(.2) in each case shows a zoomed in time window of the VLF amplitude(filtered for powerline harmonics) from both the EW and NS antenna as a function of time around the time of the impulsive VLF signal showing a consistent azimuth (the red line in the first plot). If multiple peaks are present, and the peaks are all similar, the signal of the strongest peak is shown. Here the vertical red lines denote the start and end of the fourier transform window used to analyze this signal. The plot itself shows ±2ms around the start of the signal.

The third plot(.3) shows the same information as plot #2 but with a ±10ms time window.

Finally, the fourth plot(.4) for each meteor-VLF event shows the ratio of the signal amplitude from the EW/NS antennas from the strongest peak in ~500 Hz bandpasses from 100 Hz - 20 kHz, each bandpass being represented by a single blue dot. The best fit line to these points represents the best estimate of the VLF signal arrival azimuth and is shown by a blue solid line. The azimuthal extent of the meteor as seen from Elginfield is shown as green lines. When signals from different frequency bins show different direction(obvious outliers), said outliers are removed to make line of best fits better fit the data points.

| Event Time | Magnitude | Duration of luminous flight(s) | Azimuthal Spread(degrees) | Station |
|---|---|---|---|---|
| 2017/08/23 06:25.22 | -5.65 | 3.603 | 39 | elginfield |
| 2017/09/10 03:45.56 | -5.02 | 1.268 | 10.9 | elginfield |
| 2017/09/12 08:10.29 | -4.91 | 0.501 | 48 | elginfield |
| 2017/11/15 04:50.20 | -5.51 | 0.767 | 3 | elginfield |
| 2017/12/13 09:49.50 | -5.40 | 0.934 | 16.2 | other |
| 2018/08/13 08:01.08 | -4.80 | 0.301 | 1.9 | other |
| 2018/08/15 08:56.28 | -4.80 | 0.367 | 26.46 | elginfield |

Table A1. Time/magnitude/duration/azimuthal spread/station of the seven events. Camera data from Elginfield observatory is used whenever possible, even if another stations detect a brighter magnitude compared to Elginfield. For multistation detection where Elginfield did not see the meteor, the station that recorded the most amount of frames is used, even if other stations detect a brighter magnitude. Event number 3 and 7, Elginfield detects dimmer than -5 magnitude, but other stations detect brighter than -5 magnitude so the events are included for inspection. Event number 6, station #4 detected the most frame but maximum magnitude dimmer than -5, while station #7 detects brighter than -5 magnitude but less frames, therefore light curve data from station #4 is used.

Event 1)2017/08/23 06.25.22

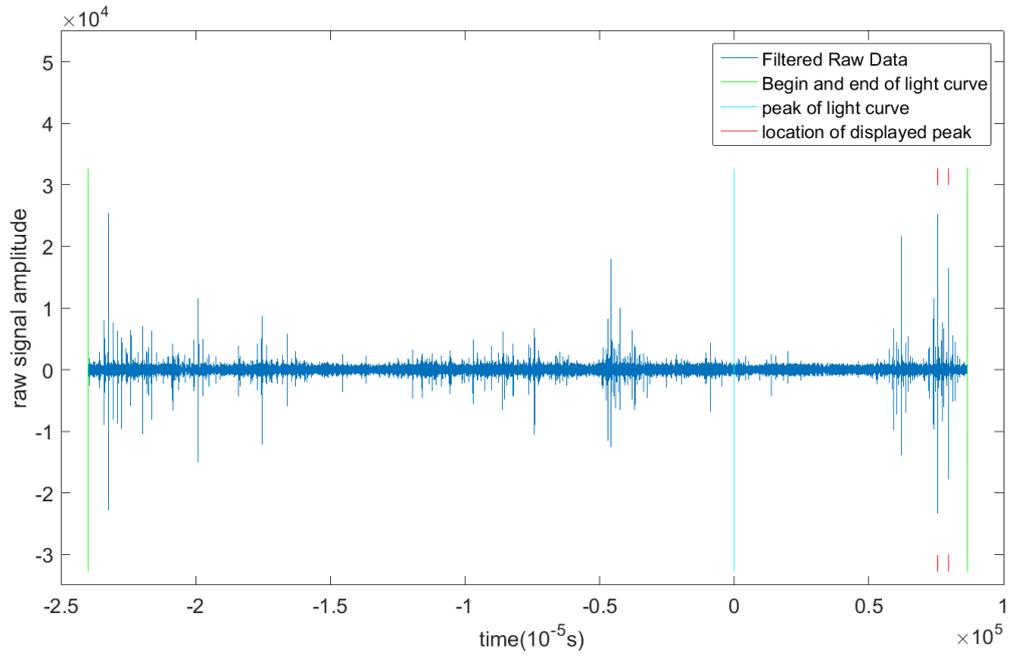

Figure A1.1) The location of 2 peaks showing matching direction

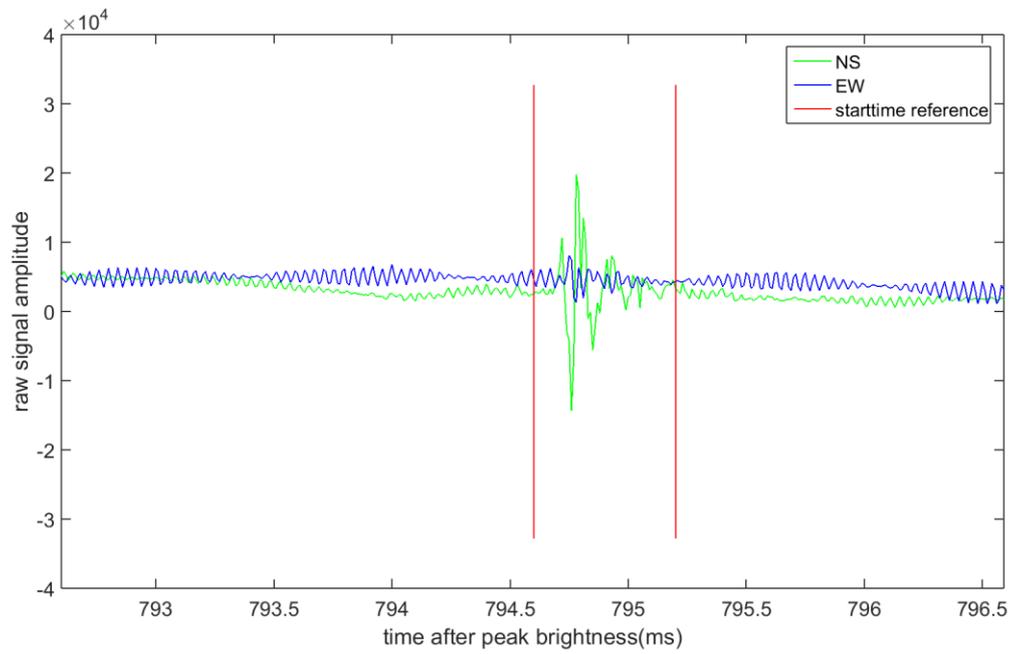

Figure A1.2) ±2ms raw amplitude of the strongest peak showing matching direction

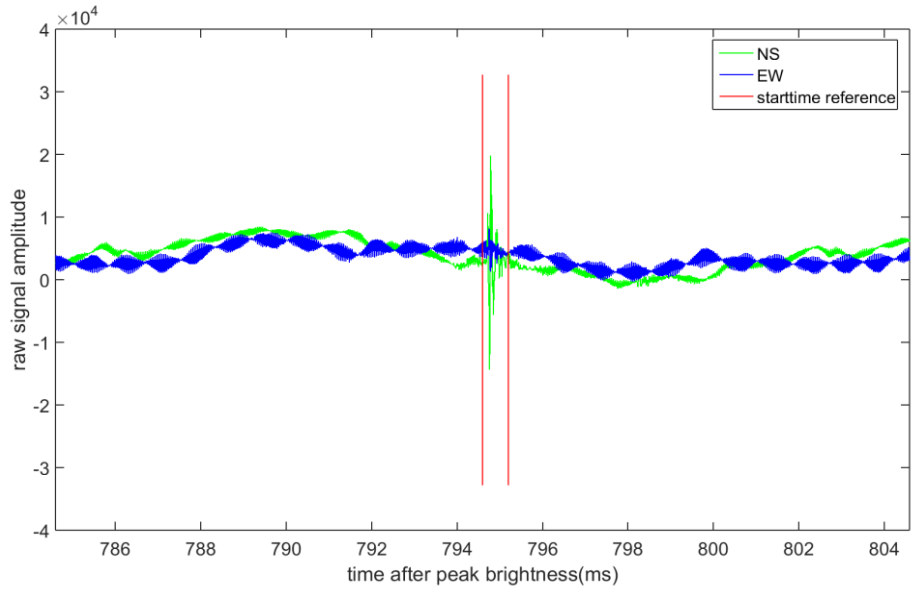

Figure A1.3) ±10ms raw amplitude of the strongest peak showing matching direction

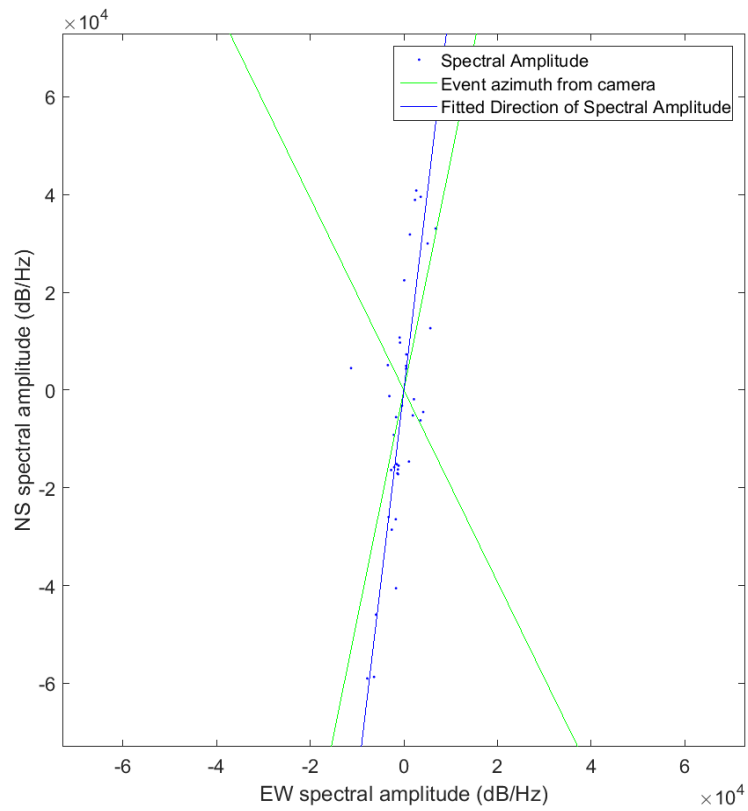

Figure A1.4) Direction fit of the strongest peak showing matching direction. All peaks show the same direction fit

Event 2) 2017/09/10 03.45.56

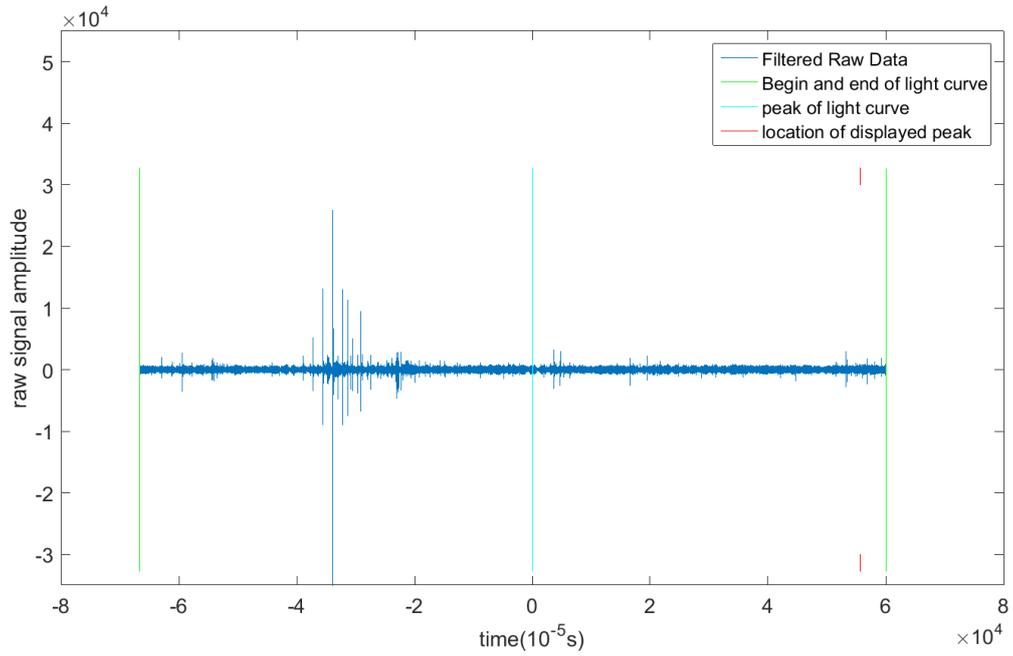

Figure A2.1) The location of the peak showing matching direction

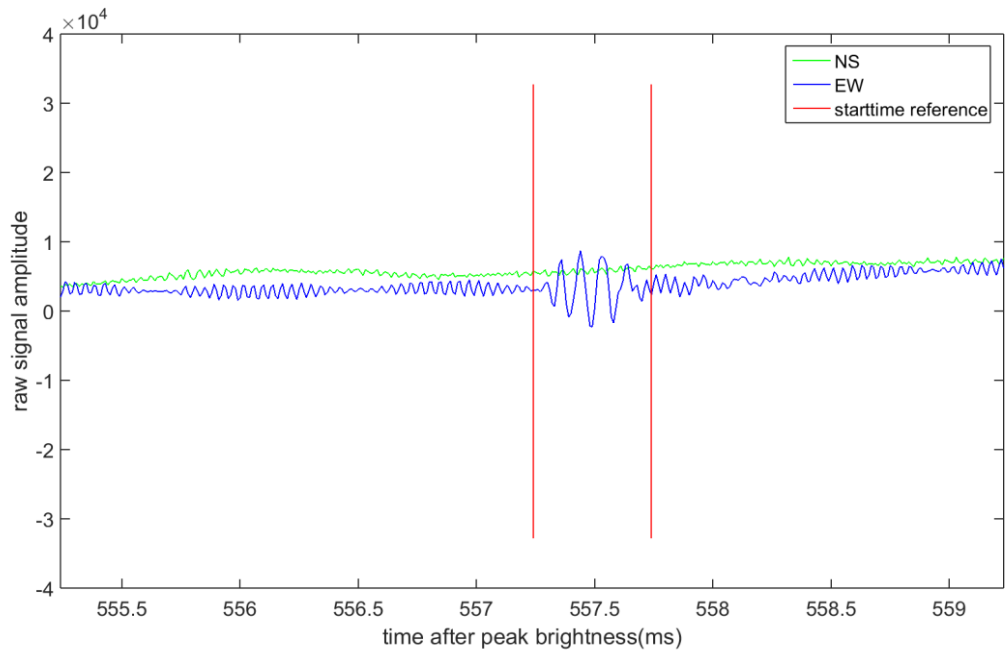

Figure A2.2) ±2ms raw amplitude of the peak showing matching direction

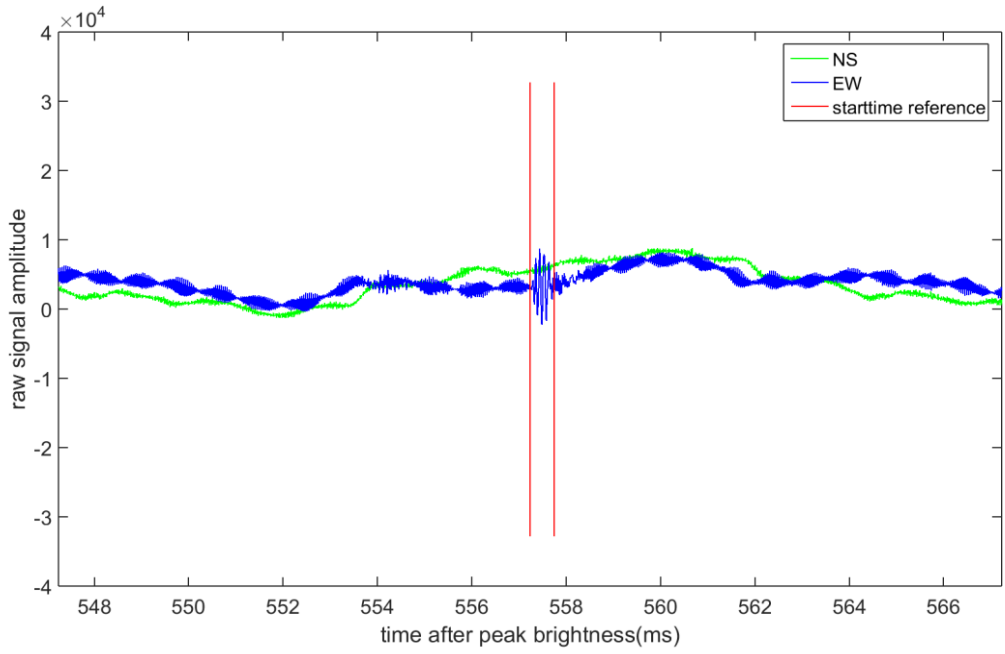

Figure A2.3) ±10ms raw amplitude of the peak showing matching direction

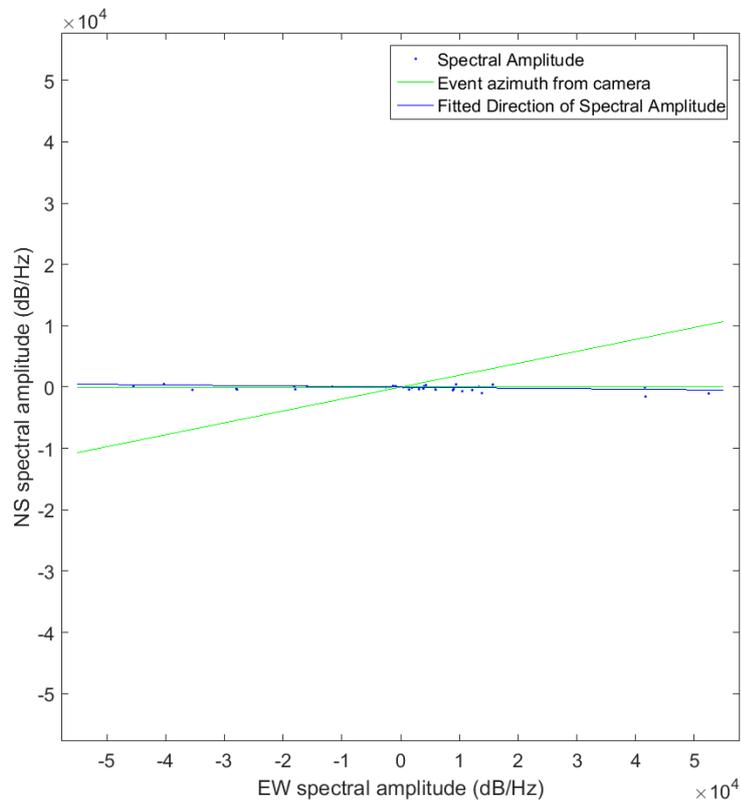

Figure A2.4) Direction fit of the peak showing matching direction.

Event 3) 2017/09/12 08.10.29

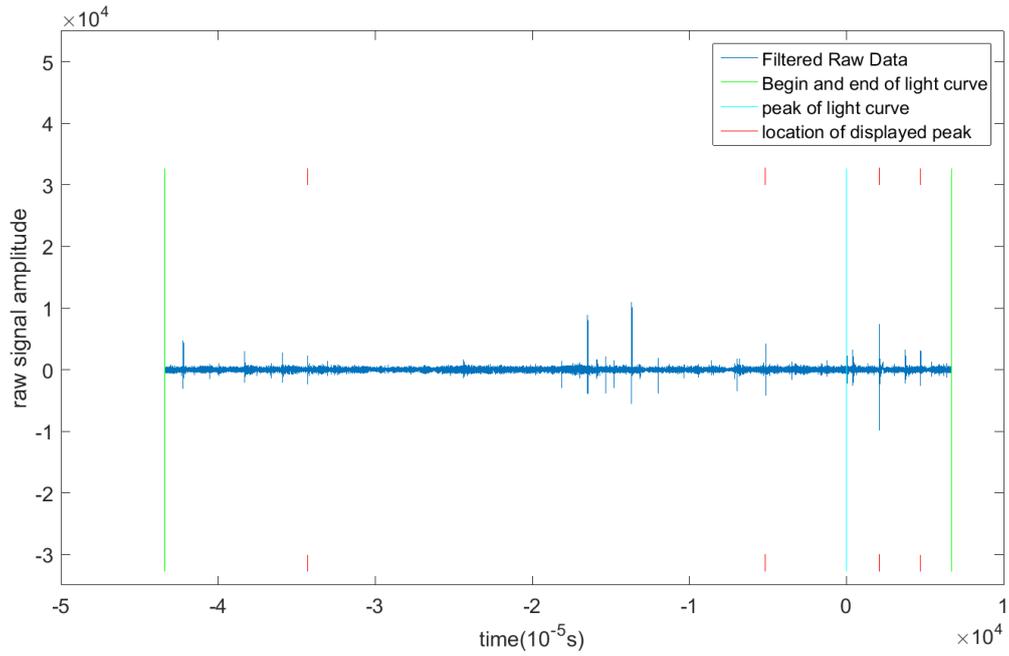

Figure A3.1) The location of the peaks showing matching direction

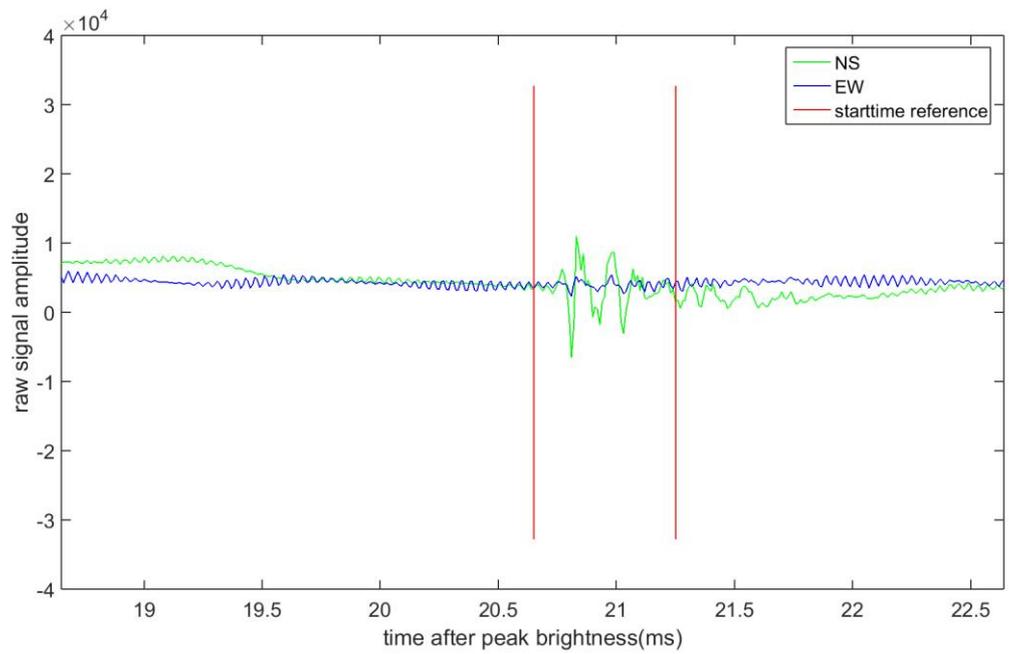

Figure A3.2) ±2ms raw amplitude of the strongest peak showing matching direction

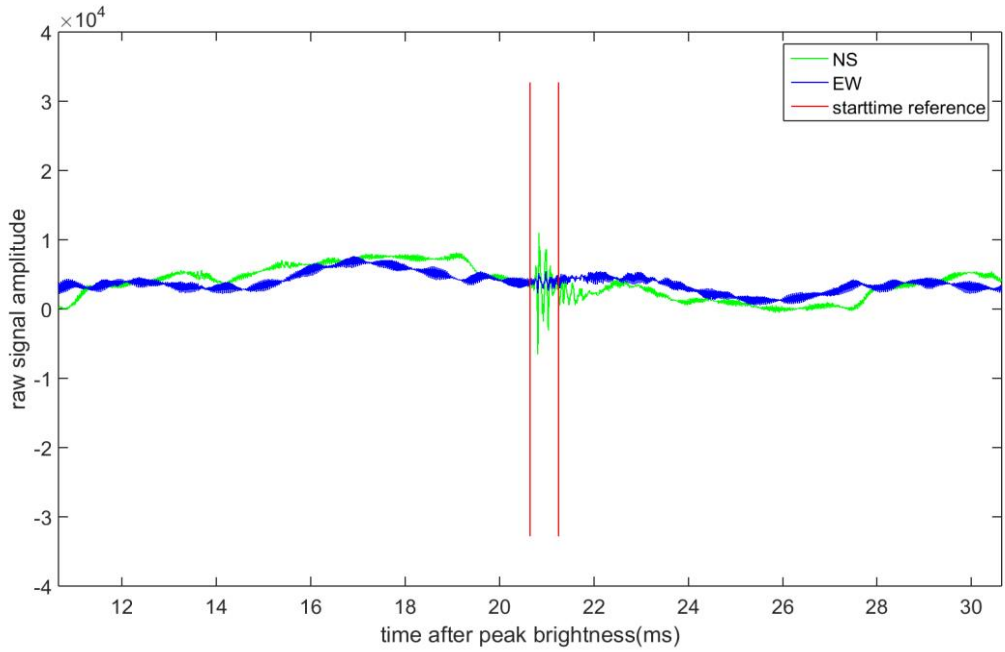

Figure A3.3) ±10ms raw amplitude of the strongest peak showing matching direction

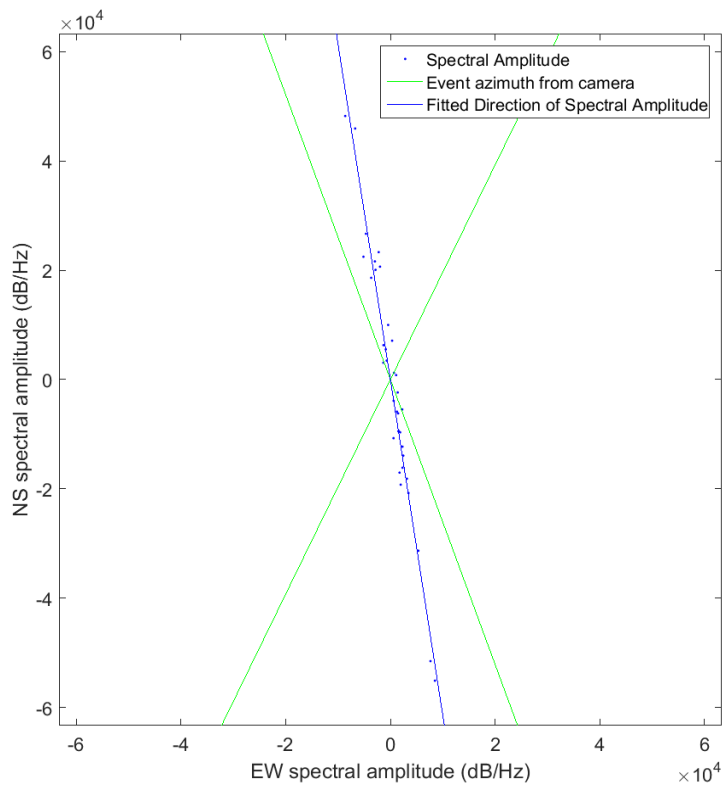

Figure A3.4) Direction fit of the strongest peak showing matching direction. All peaks show the same direction fit

Event 4)2017/11/15 04.50.20

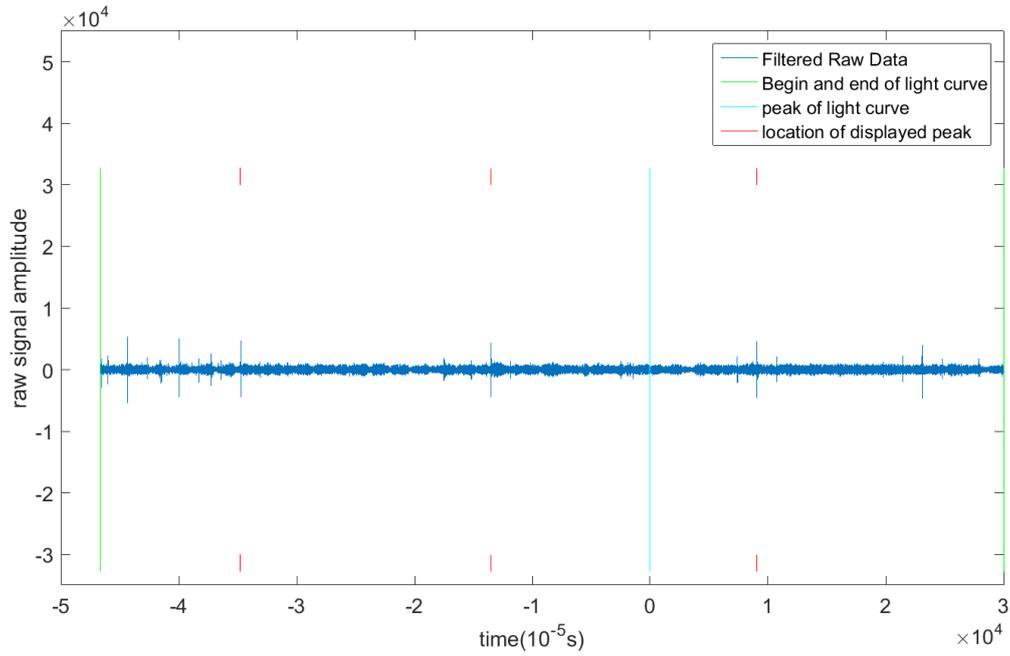

Figure A4.1) The location of the peaks showing matching direction

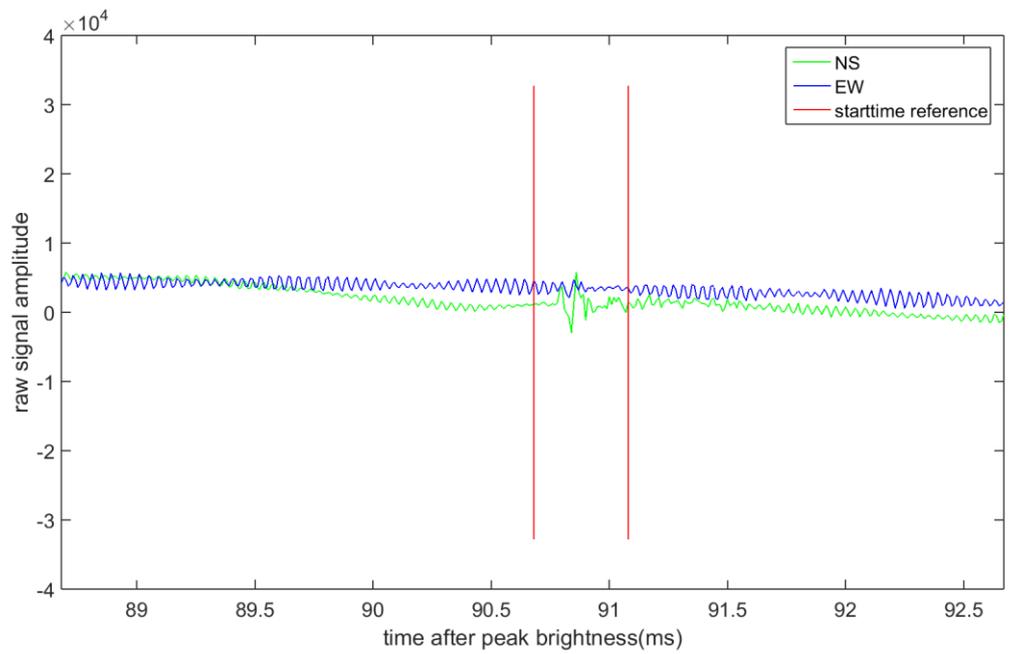

Figure A4.2) ±2ms raw amplitude of the strongest peak showing matching direction

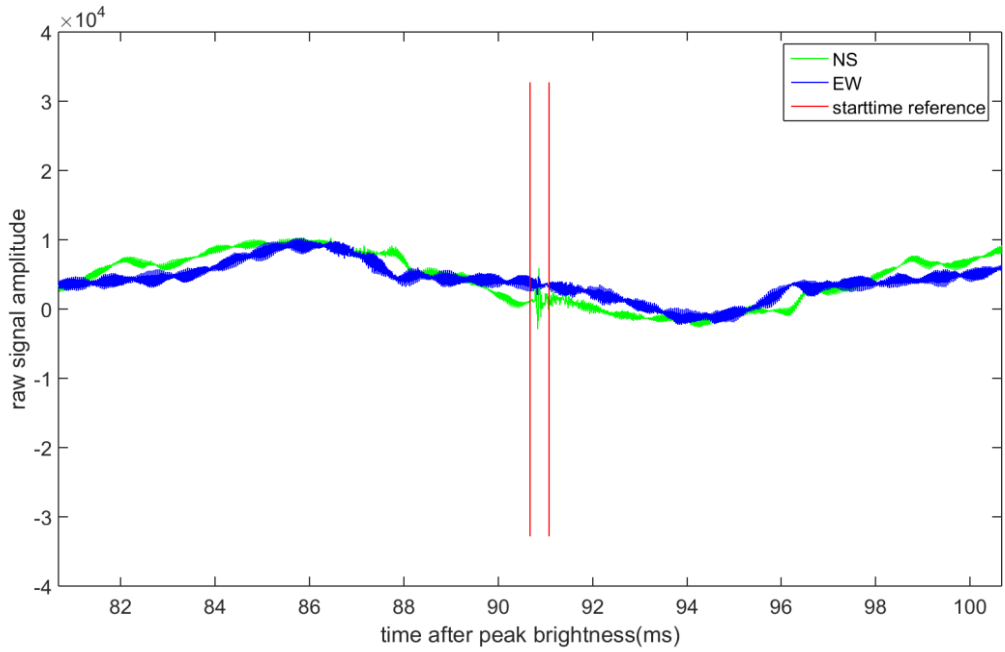

Figure A4.3) ±10ms raw amplitude of the strongest peak showing matching direction

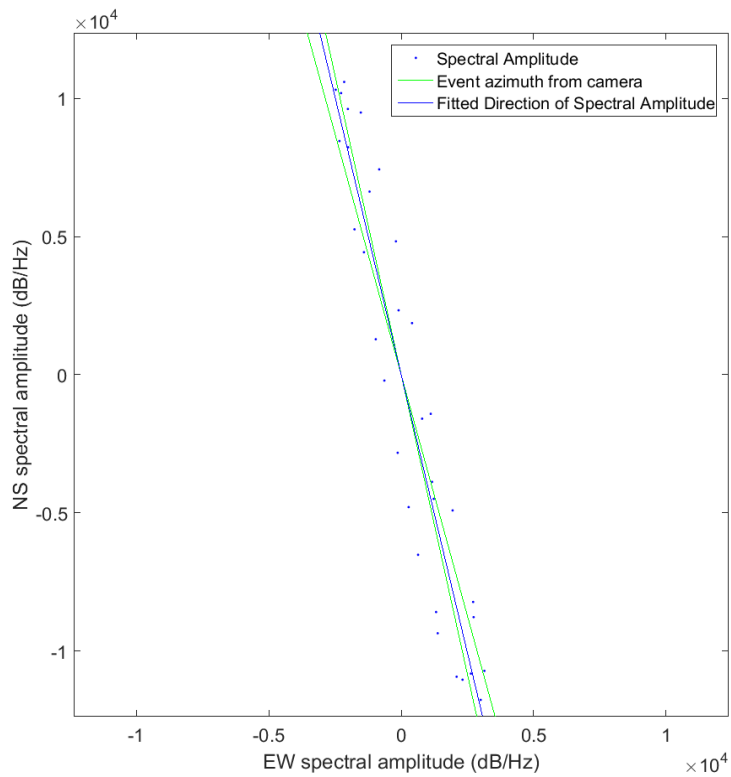

Figure A4.4) Direction fit of the strongest peak showing matching direction. All peaks show the same direction fit

Event 5) 2017/12/13 09.49.50

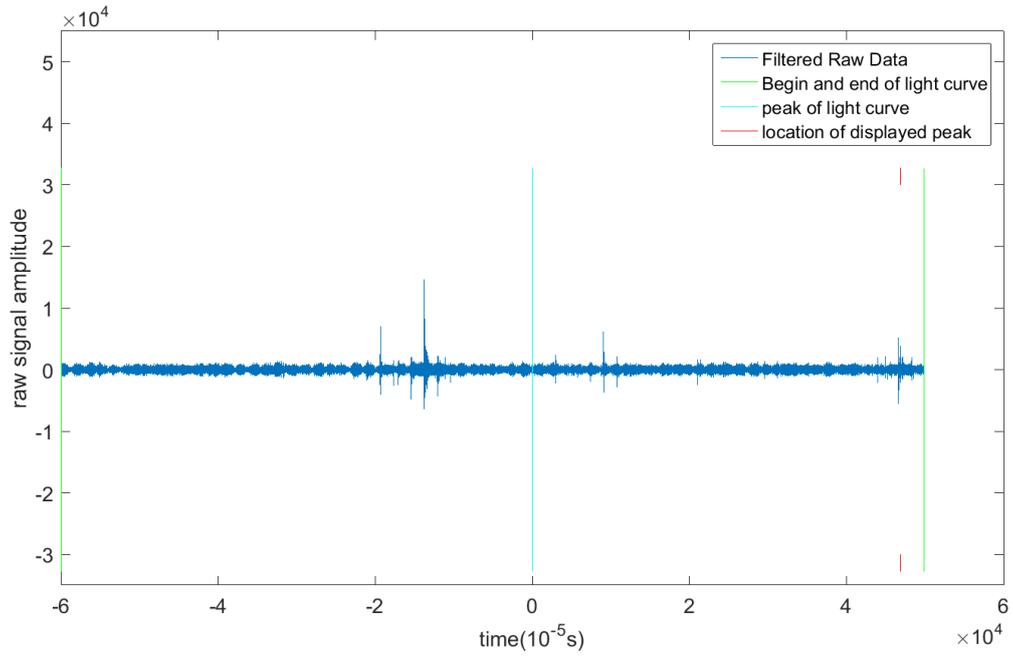

Figure A5.1) The location of the peak showing matching direction

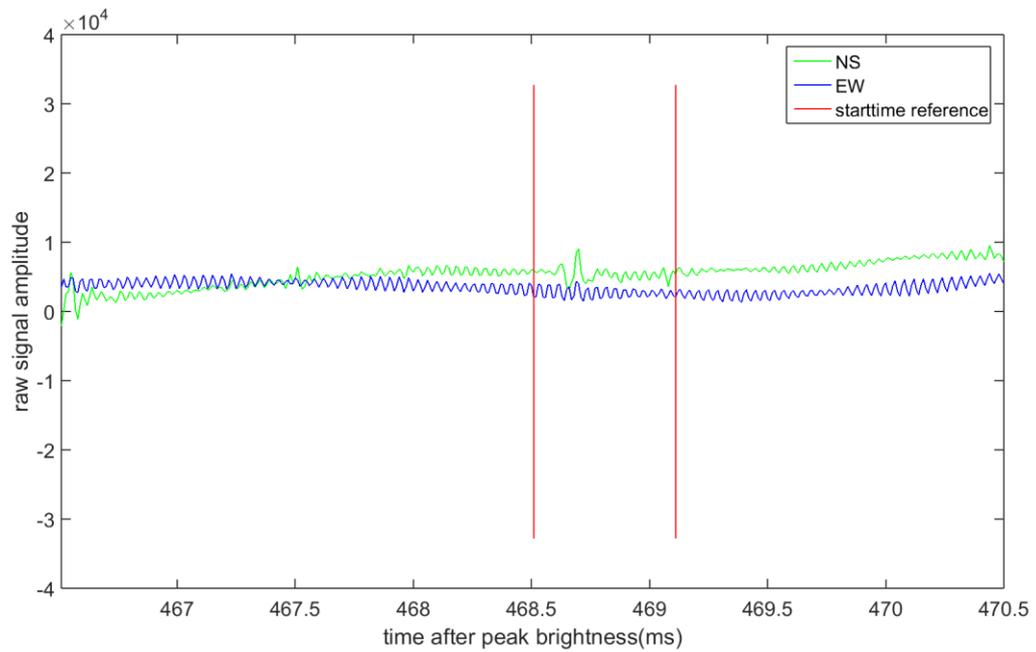

Figure A5.2) ±2ms raw amplitude of the peak showing matching direction

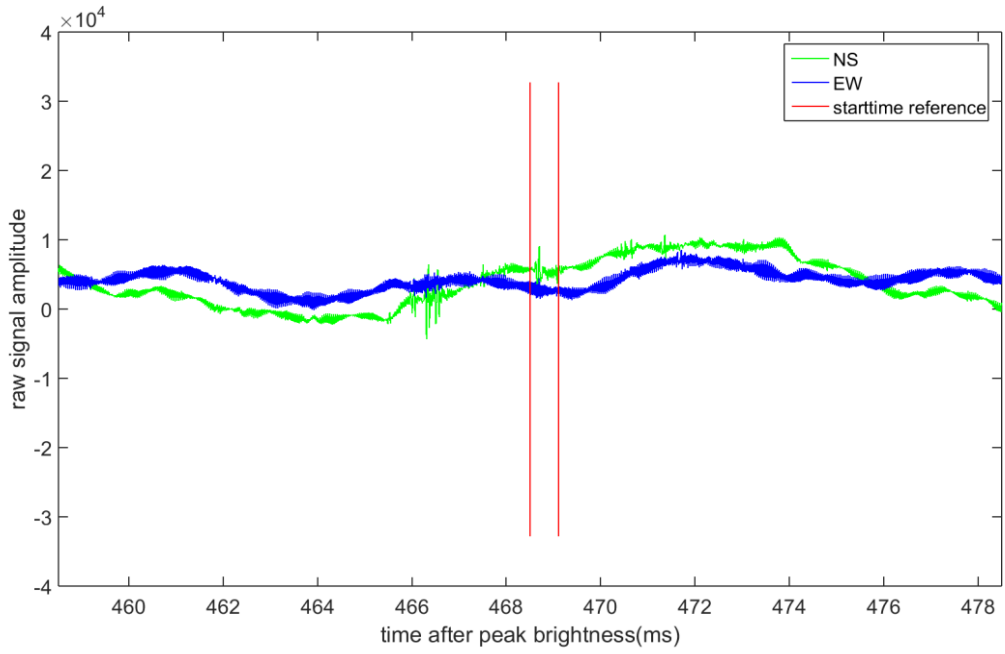

Figure A5.3) ±10ms raw amplitude of the peak showing matching direction

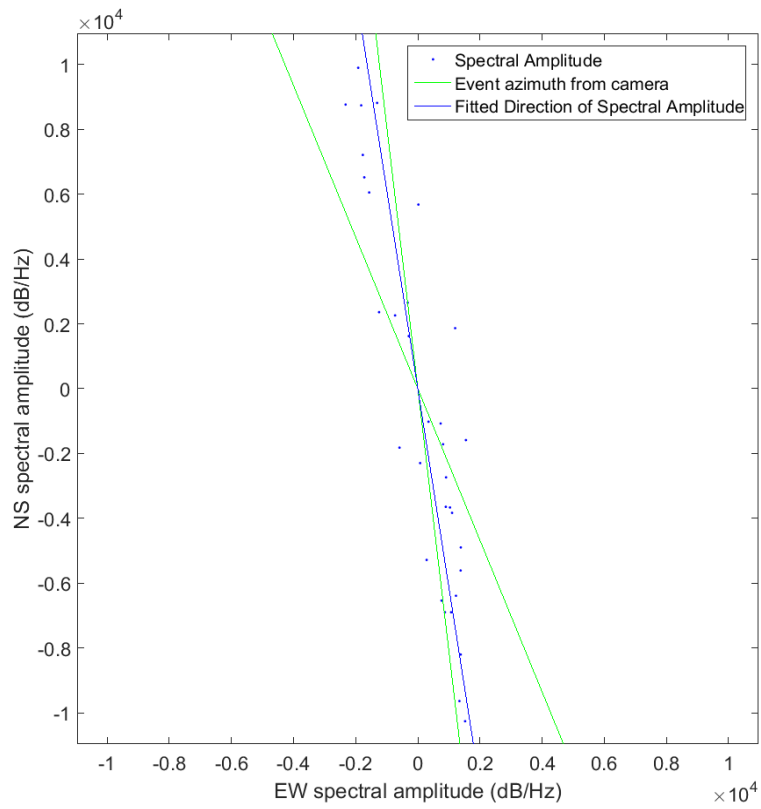

Figure A5.4) Direction fit of the peak showing matching direction.

Event 6) 2018/08/13 08.01.08

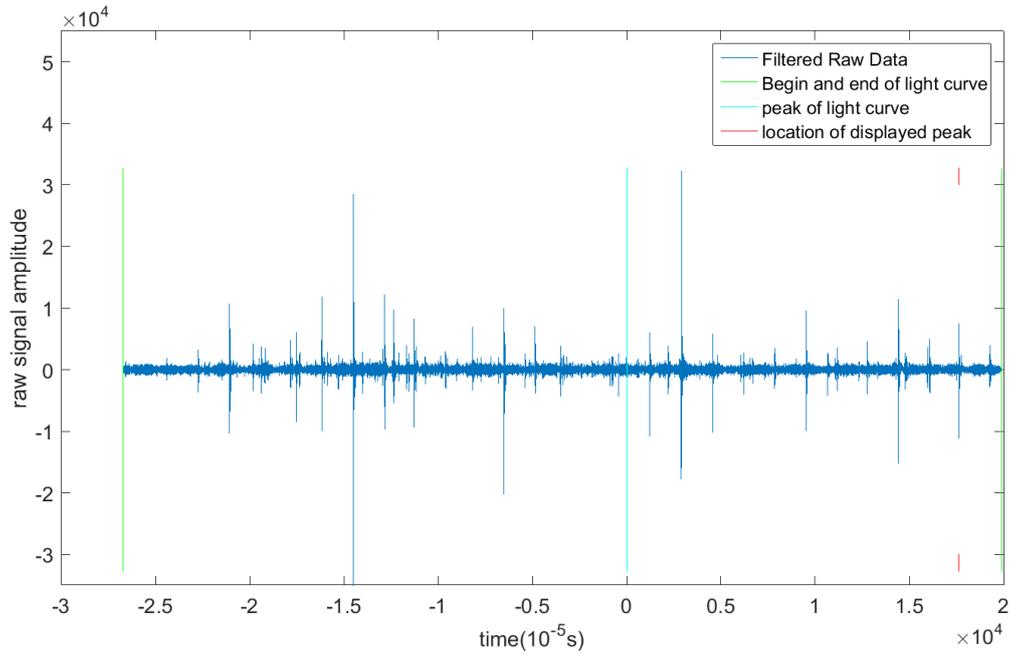

Figure A6.1) The location of the peak showing matching direction

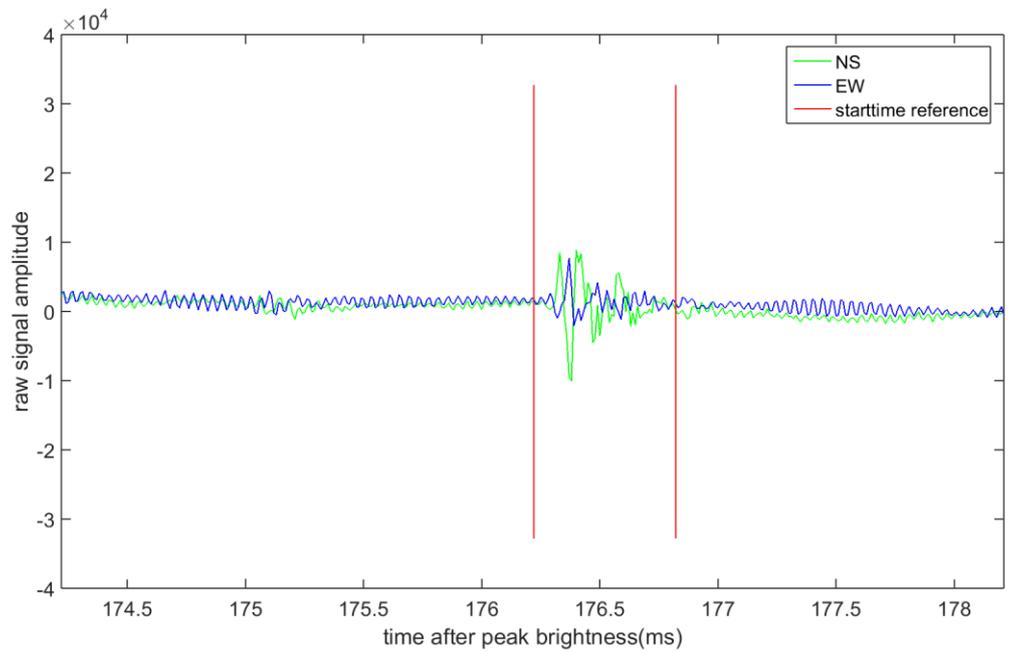

Figure A6.2) ±2ms raw amplitude of the peak showing matching direction

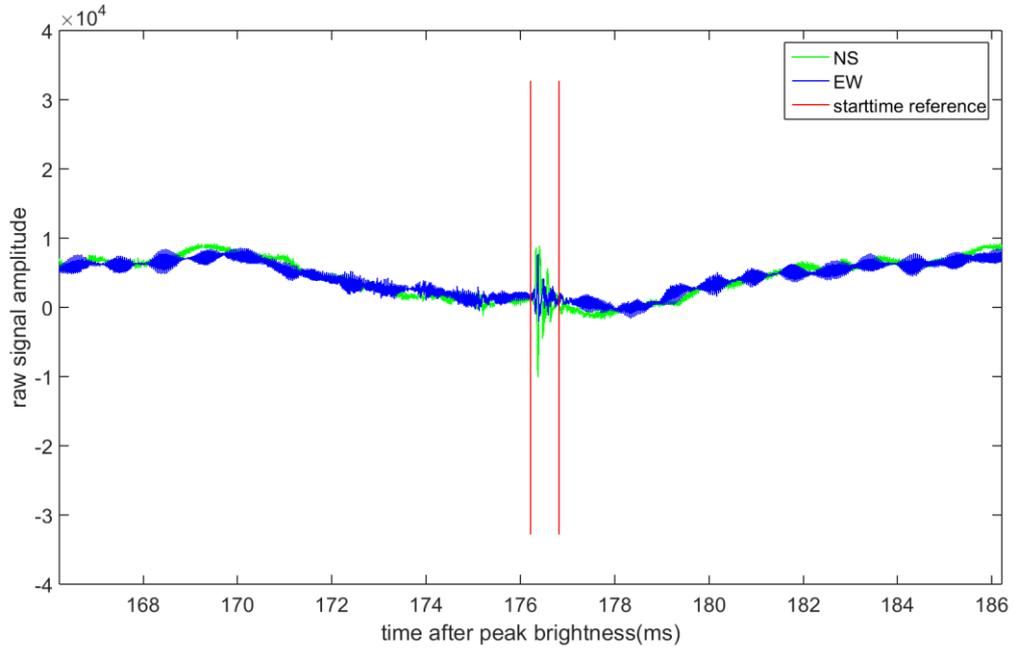

Figure A6.3) ±10ms raw amplitude of the peak showing matching direction

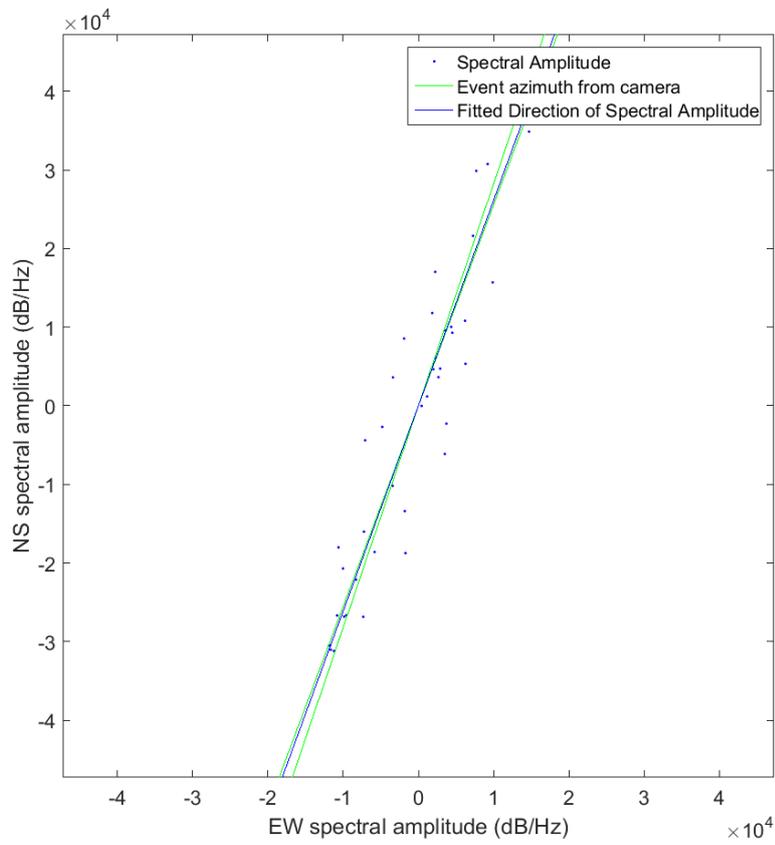

Figure A6.4) Direction fit of the peak showing matching direction.

Event 7)2018/08/15 08.56.28

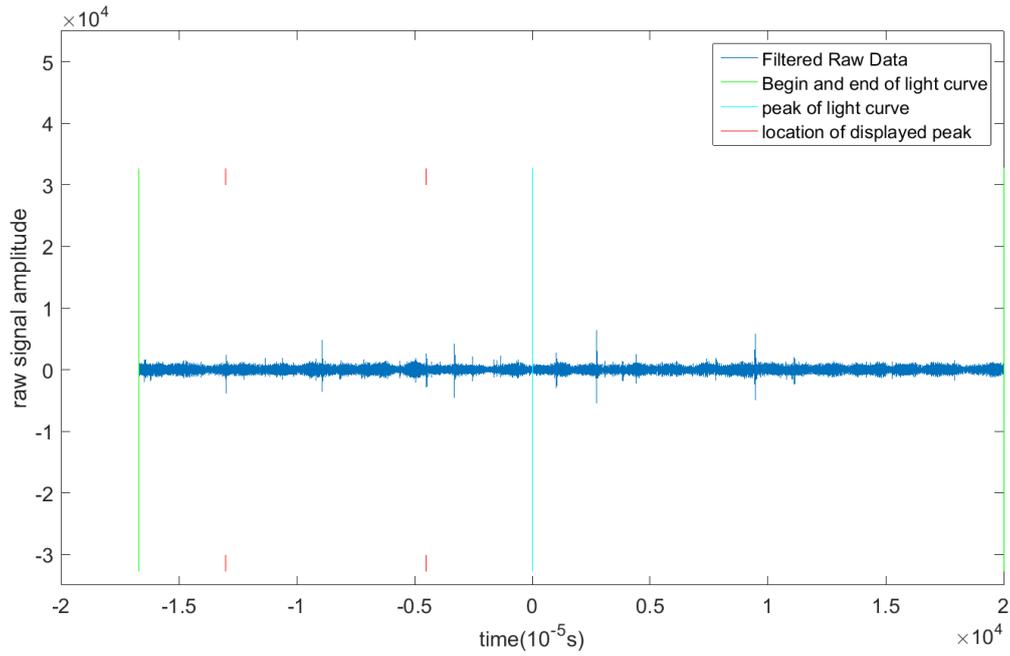

Figure A7.1) The location of 2 peaks showing matching direction

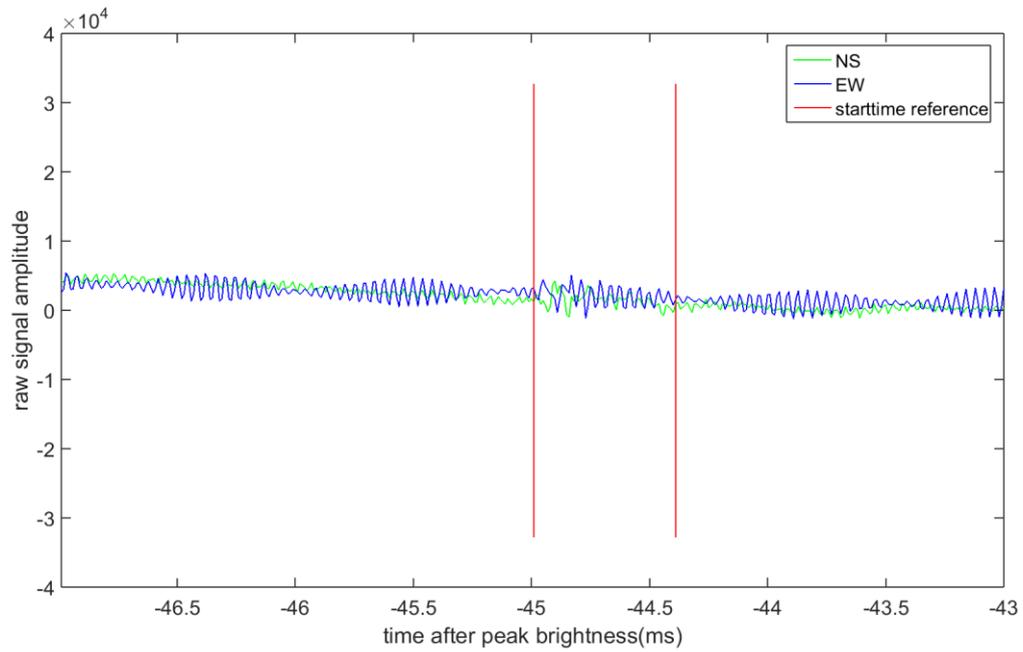

Figure A7.2) ±2ms raw amplitude of the strongest peak showing matching direction

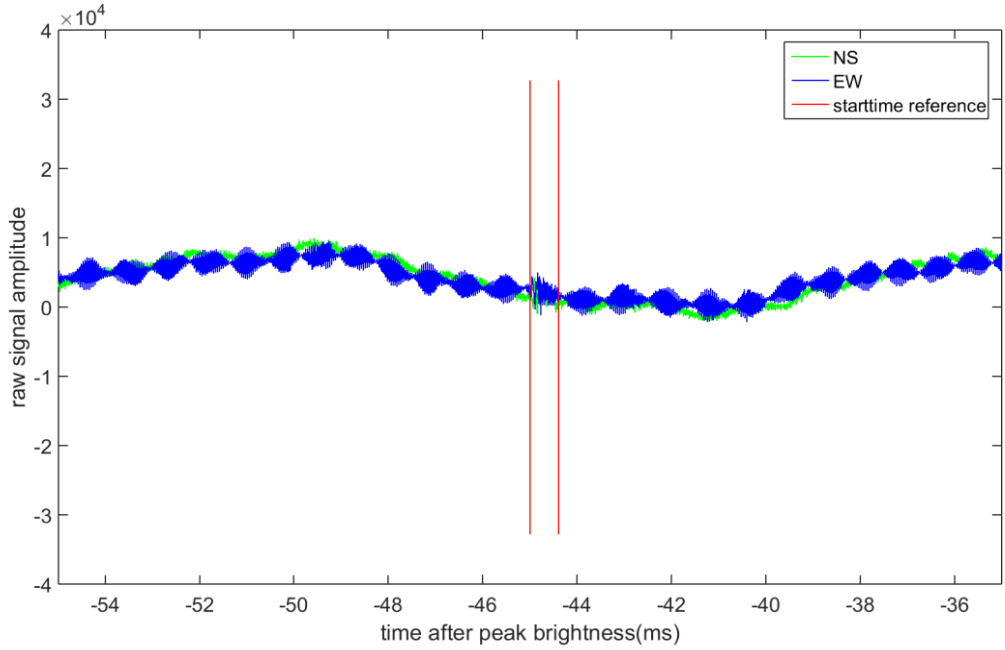

Figure A7.3) ±10ms raw amplitude of the strongest peak showing matching direction

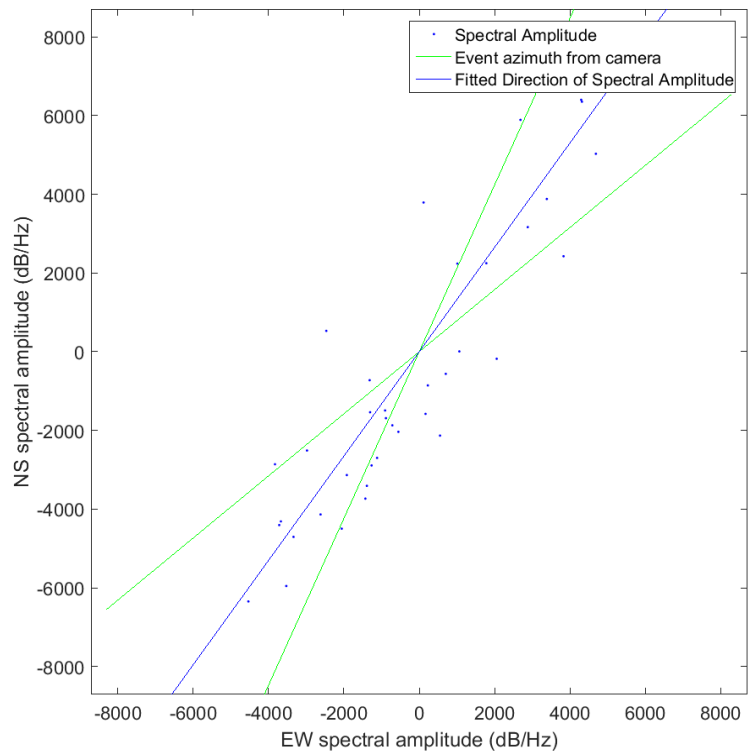

Figure A7.4) Direction fit of the strongest peak showing matching direction. All peaks show the same direction fit